\documentclass[aps,amsmath,twocolumn,floatfix,superscriptaddress,footinbib]{revtex4-1}
\pdfoutput=1
\usepackage[dvips]{graphics}
\usepackage{bm}
\usepackage{braket}
\usepackage{import}
\usepackage{epsfig}
\usepackage{enumerate}
\usepackage{subfigure}
\usepackage{color}
\usepackage{xcolor}
\usepackage{epsfig,amssymb,amsmath,latexsym}
\usepackage{sidecap,tikz}
\usepackage{graphicx}
\usepackage{makecell}
\usepackage{hyperref}
\usepackage{multirow}
\usepackage{comment} 
\usepackage{cancel}

\usepackage[normalem]{ulem}

\hypersetup{
    pdfnewwindow=true,      
    colorlinks=true,       
    linkcolor=blue,          
    citecolor=magenta,        
    filecolor=green,      
    urlcolor=purple      
}

\newcommand{\itt}{\textit}
\newcommand{\etal}{\itt{et al.~}}
\newcommand{\ie}{{i.e.~}}
\newcommand{\eg}{e.g.}
\newcommand{\bea}{\begin{eqnarray}}
\newcommand{\eea}{\end{eqnarray}}
\newcommand{\beq}{\begin{equation}}  
\newcommand{\eeq}{\end{equation}}
\newcommand{\non}{\nonumber}

\newcommand{\sg}{\sigma}
\newcommand{\eps}{\epsilon}
\newcommand{\Del}{\Delta}
\newcommand{\dis}{\displaystyle}
\newcommand{\Ps}{\Psi}
\newcommand{\UP}{\uparrow}
\newcommand{\DN}{\downarrow}
\newcommand{\K}{\mb{k}}

\newcommand{\mb}{\mathbf}

\newcommand{\mc}{\mathcal}
\newcommand{\dw}{$d$-wave }

\definecolor{lime}{HTML}{A6CE39}

\DeclareRobustCommand{\orcidicon}{\hspace{-1mm}
	\begin{tikzpicture}
	\draw[lime, fill=lime] (0,0) 
	circle [radius=0.16] 
	node[white] {{\fontfamily{qag}\selectfont \tiny \,ID}};
	\draw[white, fill=white] (-0.0525,0.095) 
	circle [radius=0.007];
	\end{tikzpicture}
	\hspace{-3mm}
}
\foreach \x in {A, ..., Z}{\expandafter\xdef\csname orcid\x\endcsname{\noexpand\href{https://orcid.org/\csname orcidauthor\x\endcsname}
			{\noexpand\orcidicon}}
}

\begin{document} 
\title{Transport signatures of Bogoliubov Fermi surfaces in normal metal/time-reversal symmetry broken 
\dw superconductor junctions}
\author{Amartya Pal\orcidA{}}
\email{amartya.pal@iopb.res.in}
\affiliation{Institute of Physics, Sachivalaya Marg, Bhubaneswar-751005, India}
\affiliation{Homi Bhabha National Institute, Training School Complex, Anushakti Nagar, Mumbai 400094, India}

\author{Arijit Saha\orcidB{}}
\email{arijit@iopb.res.in}
\affiliation{Institute of Physics, Sachivalaya Marg, Bhubaneswar-751005, India}
\affiliation{Homi Bhabha National Institute, Training School Complex, Anushakti Nagar, Mumbai 400094, India}

\author{Paramita Dutta\orcidC{}}
\email{paramita@prl.res.in}
\affiliation{Theoretical Physics Division, Physical Research Laboratory, Navrangpura, Ahmedabad-380009, India}

\date \today

\begin{abstract}
In recent times, Bogoliubov Fermi surfaces (BFSs) in superconductors (SCs) have drawn significant attention due to a substantial population of Bogoliubov quasiparticles (BQPs) together with Cooper pairs (CPs) in them. The BQPs as zero energy excitations give rise to captivating and intricate charge dynamics within the BFSs. In this theoretical study, we propose to reveal the unique signatures of the topologically protected BFSs in bulk $d$-wave SCs using normal metal/time-reversal symmetry (TRS) broken \dw SC hybrid setup, in terms of the differential conductance and Fano factor (FF). Orientation of crystal $a$ axis with respect to junction normal, quantified by the parameter $\alpha$, is crucial for transport properties in these hybrid devices. For $\alpha=0$, an enhancement in zero-bias conductance (ZBC) can be identified as a key signature of BFSs. However, for $\alpha\ne0$, this feature does not replicate due to the presence of the localized Andreev bound state (ABS) at the interface. The interplay of ABS and BFSs gives rise to an anomalous behavior in ZBC compared to the $\alpha=0$ case. This behavior remains qualitatively similar even at finite temperatures. Finally, we explain this anomalous behavior by analyzing the effective charge of the carriers in terms of the FF. The simplicity of our setup based on \dw SC makes our proposal persuasive.
\end{abstract}


\maketitle


\section{Introduction}
The appearance of the gap in the Bogoliubov quasiparticle (BQP) spectrum is a characteristic property of conventional Bardeen-Cooper-Schrieffer (BCS) SCs\,\cite{Bardeen1957}. For many unconventional SCs, this gap in the momentum space vanishes either at specific point nodes or along line nodes whose dimensions are always less than the dimension of the underlying normal state Fermi surface (FS)\,\cite{Sigrist1991}. However, the density of BQPs at these nodes appear to be very low. Very recently, FS with same dimensions as that of the underlying normal state FS and substantially enhanced density of states, have been theoretically proposed by Agterberg \etal\,\cite{Agterberg2017}. Interestingly, these FSs known as Bogoliubov Fermi surfacess (BFSs), are topologically protected by certain combinations of discrete symmetries and characterized by a $\mathcal{Z}_2$ topological invariant. 

In the literature, the concept of BFS was first proposed in time-reversal symmetry (TRS)-broken and inversion-symmetric multiband SCs with total angular momentum $j=3/2$~\cite{Agterberg2017}. From several followup works, the general prescription for the appearance of BFSs is provided as: SCs 
possessing additional degrees of freedom (e.g.\,orbital or sublattice) other than spin, host both interband and intraband pairings, out of which a pseudomagnetic field appears and causes BFSs\,\cite{Agterberg2017,Brydon2018,Menke2019,Sumita2019,Lapp2020,Tamura2020,Oh2021,Kim2021,Dutta2021,Zhu2021,Bhattacharya2023,Miki2023}. For such exotic nature of pairing states, iron-based superconductors~\cite{Gao2010,Nicholson2012,Ong2016,Vafek2017,Agterberg2017b,Hu2020}, half-Huesler compounds~\cite{Agterberg2017,Brydon2018,Brydon2016,Savary2017,Yang2017,Roy2019,Timm2017,Boettcher2018,Boettcher2018b}, UPt$_3$~\cite{Yanase2016,Nomoto2016}, twisted bilayer graphene~\cite{Guo2018,Su2018,Wu2019} etc. are proposed as candidate materials. Furthermore, BFSs are also shown to exist together with odd-frequency pairing in $j=3/2$ SCs\,\cite{Dutta2021}. The presence of inversion symmetry is crucial for the $\mc{Z}_2$ invariant\,{\cite{Brydon2018,Timm2021a,Agterberg2017,Bzdu2017}.  By contrast, the existence of BFSs is also investigated in non-centrosymmetric SCs~\cite{Timm2021b,Timm2017,Link2020b} since inversion-symmetry can bring instability to BFSs under generic attractive interactions between BQPs\,\cite{Herbut2021,Oh2020,Tamura2020,Link2020a}. 
In addition to $j=3/2$ SCs, BFSs are also proposed to exist in TRS-broken spin-$1/2$ SCs\,\cite{Setty2020PRB,Cao2023,Banerjee2022}. Explicitly, SCs with interband pairing~\cite{Setty2020NatComm} and with large supercurrents\,\cite{Fulde1965,Setty2020PRB} are shown to host BFSs.

BFSs are predicted to observe in Sr$_2$RuO$_4$~\cite{Suh2020}, URu$_2$Si$_2$~\cite{Kasahara2007}, UTe$_2$~\cite{Ran2019,Metz2019} etc. with reports on residual density of states (DOS) which is considered to be one of the prime signatures of BFSs. The possibility of experimental realization of BFSs in a multiband model by measuring electronic specific heat, tunneling conductance, thermal conductivity, magnetic penetrations depth, NMR spin-lattice relaxation and spin-lattice relaxation rate etc. have been discussed earlier in the literature~\cite{Lapp2020,Setty2020PRB}. Very recently, segmented BFSs have been discovered in Bi$_2$Te$_3$ thin films in proximity with NbSe$_2$ SC\,\cite{Zhu2021} whereas, the previous proposals involve bare bulk SCs. The emergence of BFSs in hybrid junctions of $s$-wave SCs is further supported by the observation in Al-InAs junction in the presence of magnetic field\,\cite{Phan2022}. These recent discoveries spur further research activities on BFSs in SC hybrid structures. Till date, there is only one theoretical work on BFSs in heterostructure where normal metal/$s$-wave SC junctions are considered in presence of in-plane Zeeman field and Rashba spin-orbit coupling~\cite{Banerjee2022}. These works invoke further questions about the possibility of detecting the signatures of BFSs in $d$-wave SCs since most of the earlier proposals are based on bulk $d$-wave pairing\,\cite{Setty2020PRB,Setty2020NatComm,Link2020a,Yang1998,Dutta2021,Christos2023}. 

In this article, we propose to capture the signatures of BFSs in bulk $d$-wave SCs via possible experimental observables using normal metal/$d$-wave SC heterostructure. The coexistence of BQPs and Cooper pairs (CPs) in the BFS provides a rich playground for studying transport properties and the Fano factor (FF) characteristics in a simple setup. We show that topologically protected BFSs appear in bulk \dw SC subjected to an in-plane Zeeman field. Employing Blonder-Tinkham-Klapwijk (BTK) formalism, we calculate the differential conductance by forming a metal/SC hybrid junction (as shown in Fig.\,\ref{fig:Schematic_NS}($a$)) and find an enhancement in zero-bias conductance (ZBC), providing a key signature of BFSs. We extend our study to nonzero values of $\alpha$ to explore the effect of anisotropy on BFSs. The emergence of zero-energy interface-localized Andreev bound states (ABS) is the previously established classic signature of anisotropy in \dw SCs\,\cite{Hu1994,Nagato1995,Tanaka2021,Tamura2017}. The interplay between ABS and BFSs gives rise to intriguing features of ZBC for the anisotropic case. We also analyse the effect of finite temperature on differential conductance of our system. To confirm the role of BFSs in the anomalous behavior of the conductance, we perform a detailed analysis of shot-noise spectroscopy and FF therein. We use both continuum and lattice model, and find a perfect match between the results obtained in the two models, establishing the robustness of the transport signatures of BFSs. 

The remainder of the paper is organized as follows. In Sec.~\ref{sec:II}, we describe our model and topological characterization of BFSs. Sec.~\ref{sec:III} and Sec.~\ref{sec:IV} are devoted 
to the discussion of transport and FF signatures of BFSs in our setup.  Finally, we summarize and conclude the paper in Sec.~\ref{sec:V}.

\begin{figure}[!h]
	\centering
	\includegraphics[scale=0.305]{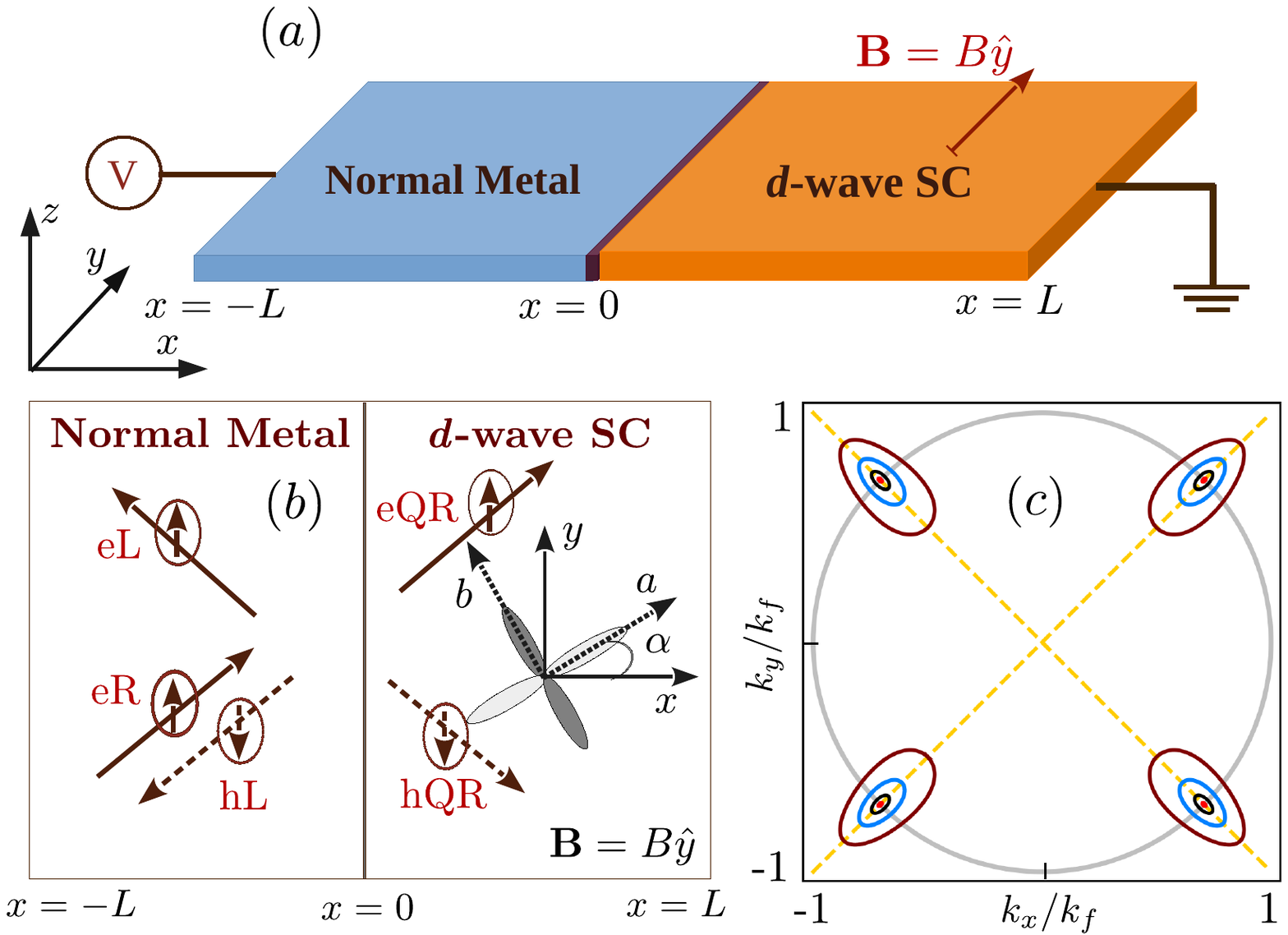}
\caption{($a$) Schematic diagram of normal metal/\dw SC heterostructure with an externally applied Zeeman field $\mathbf{B}$ and voltage bias $V$. $(b)$ The scattering processes at the normal metal/\dw SC junction is depicted where an incident up-spin ($\uparrow$) right-moving electron (eR) at the interface ($x$=0) can be scattered either as a normally reflected (left-moving) $\uparrow$ electron (eL) or Andreev reflected down-spin ($\downarrow$) hole hL, or transmitted as $\uparrow$ ($\downarrow$) electron (hole)-like quasiparticles eQR (hQR). Here, $\alpha$ is the angle between crystal $a$-axis and the interface normal.  $(c)$ The emergence of four BFSs with the Zeeman field $B=0.1$ (black), $0.2$ (blue), $0.5$ (brown). The normal state FS, four point nodes, and two line nodes are shown by the grey circle, red dots, and yellow dashed lines, respectively.}
	\label{fig:Schematic_NS}
\end{figure}

\section{Model $\&$ topological BFS}\label{sec:II}
We consider a junction comprising of a normal metal and \dw SC (N/$d$-wave SC) in presence of a Zeeman field (only in the SC side \ie $x>0$) applied along $\hat{y}$ direction as shown in Fig.\,\ref{fig:Schematic_NS}(a). We write the Bogoliubov-de Gennes (BdG) Hamiltonian: $H_{{\rm{BdG}}}= \frac{1}{2} \sum_k \Psi_{\mb{k}}^\dagger\mc{H}(\mb{k})\Psi_{\mb{k}} $ in Nambu basis as
$\Ps_\mb{k}=[c_{\mb{k},\UP},c_{\mb{k},\DN},c^\dagger_{-\mb{k},\DN},-c^\dagger_{-\mb{k},\UP}]^T$,  with $c_{\mb{k}\sg}(c_{\mb{k}\sg}^\dagger)$ denotes annhilation (creation) operator for electrons with momentum $\mb{k}$ ($=\{k_x,k_y\}$) and spin $\sg$. We describe $\mc{H}(\mb{k})$ as\,
\cite{Setty2020PRB,Yang1998},
\beq
\mc{H}(\mb{k}) =\eps(\mb{k}) \pi_z  \sg_0 -B \pi_0  \sg_y + \Del(\mb{k}) \pi_x \sg_0\ . 
\label{Eq:ham}
\eeq
The Pauli matrices $\mb{\pi}$ and $\mb{\sg}$ act on the particle-hole and spin degrees of freedom, respectively. The kinetic energy is given by: $\dis{\eps(\mb{k}) = \frac{\mb{k}^2}{2m}}-\mu$ where 
$m$ is the effective mass of electrons and $\mu$ is the chemical potential. Also, $B$ denotes the strength of external in-plane Zeeman field. The superconducting pair potential is considered as, $ \Del(\mb{k}) = \Del_0 \cos[2(\theta + \alpha )]$ with $\theta=\tan^{-1}(k_y/k_x)$ and $\alpha$ denotes the angle 
between $a$ axis of the crystal and normal to the interface [see Fig.\,\ref{fig:Schematic_NS}(b)]. Throughout our study, we consider natural unit, explicitly, $m=\Delta_0=\hbar=1$, and set $\mu=10\Del_0$, $\alpha=0$, and $\alpha=\pi/4$. Specifically, for $\alpha=0$ ($d_{x^2-y^2}$ pairing) and $\alpha=\pi/4$ ($d_{xy}$ pairing), the pairing potential takes the following form: $\Delta_{\mb{k},\alpha=0}=\Delta_0 (k_x^2 -k_y^2)/k_f^2$ and $\Delta_{\mb{k},\alpha=\pi/4}=2\Delta_0 k_x k_y/k_f^2$, respectively, where $k_f=\sqrt{2m\mu}$ is the Fermi momentum \cite{Hu1994,Tanaka2021,Zhu1999}.  The external magnetic field is applied only in the superconducting side of the junction ($x>0$) and only couples with the spin degree of freedom of electrons.There is no coupling of external magnetic field to the orbital degree of freedom for simplicity; thus no orbital magnetization present in the system. The change in other parameter values do not qualitatively affect the main message of the present work. Note that, $\Delta_{0}=0$ in the normal part of the junction ($x<0$).

Before we proceed calculate the two-terminal conductance and the Fano factor (FF) using the geometry shown in Fig.\,\ref{fig:Schematic_NS}(a), we investigate the nature of Fermi surfaces and the density of states (DOS) considering a bare $d$-wave SC with periodic boundary conditions along $x$ and $y$ directions. In the presence of an external in-plane magnetic field,
four BFSs appear in the momentum space centred around $\mb{k}=(\pm k_f,\pm k_f)/\sqrt{2}$. They are illustrated by the zero-energy contours in Fig.~\ref{fig:Schematic_NS}(c). The area of each contour continuously increases as we enhance the strength of the Zeeman field. The topological property of BFSs can be identified by calculating the Pfaffian\,\cite{Agterberg2017,Brydon2018}. For the present model, it takes the form - $Pf(\K)= \eps(\K)^2 + \Del(\K)^2 - B^2$, similar to the model considered by Setty et al.\,\cite{Setty2020PRB}, and $Pf(\K)$ changes its sign accross each closed contour of BFS. To quantify it, a $\mc{Z}_2$ topological invariant, $\nu$, is defined as\,\cite{Agterberg2017,Brydon2018}: $(-1)^\nu = sgn[Pf(\K) Pf(0)]$, where, $Pf(0)$ is computed at the origin in the momentum space. The closed contours of BFSs separate $\nu=0$ (outside) from the $\nu=1$ (inside) regions, indicating the topological protection of BFSs in our model. We refer to the Appendix~\ref{SM:Sec_1} for more details on BFSs, and Pfaffian calculation.


With the understanding of the topological property of BFSs, we now focus on the four quantum mechanical scattering processes taking place at the N/SC junction as schematically shown in Fig.\,\ref{fig:Schematic_NS}(b). A right-moving electron with spin $\sg$ can (i) reflect as an electron with spin $\sigma$, called ordinary or normal reflection (NR), (ii) combine with another electron of opposite spin $\bar{\sigma}$ to form a spin-singlet CP leaving behind a hole to reflect from the interface due to the Andreev reflection (AR) process (for sub-gap energy), transmit as (iii) electron-like quasiparticle (QP) or (iv) hole-like QP. 

\section{Differential conductance}\label{sec:III}
In order to find the signature(s) of BFSs, we calculate the differential conductance with respect to the voltage bias $V$ applied across the junction at finite temperature $T$ employing the BTK formula\,\cite{Blonder1982,Dumitrescu2015}
%
%

\begin{equation}
	G(eV,T)\! =\! \frac{e}{h} \!\int\limits_{-\infty}^{\infty}\!\!\!\!\! dE \, \mc{T}(E,T) \frac{\partial}{\partial V}\!\left[ f_L(E,\mu_L,\!T) \! -\!\! f_R(E,\mu_R,\!T)\right]  \label{Eq:BTK_formula}
\end{equation} 
where, $\mc{T}(E,T)=\sum_{\sg=\UP,\DN}\int_{-\pi/2}^{\pi/2}d\theta_e \cos\theta_e [1-\mc{R}^e_\sg(\theta_e) + \mc{R}^h_{\sg}(\theta_e)]$. $\mc{R}^e_\sg\!=\!\! |r^{ee}_{\sg}|^2 \! $ ($\mc{R}^h_{\sg}\!\!=\!\frac{\cos\theta_h}{\cos\theta_e}|r^{eh}_{\sg}|^2)$ denotes the probability with $r^{ee}_{\sg}$ ($r^{eh}_{\sg})$ being the amplitude for the NR (AR) from the interface, respectively. Also, $\theta_e$ ($\theta_h$) is the angle of incidence (reflection) of electron (hole). Here, $f_L(E,\mu_L,\!T) (f_R(E,\mu_R,\!T))$ represents the Fermi distribution function in the left (right) sectors of our setup fixed at chemical potential, $\mu_L=eV~(\mu_R=0)$. The superconducting pair potential takes the form~\cite{Enoksen2012,Tao2012,Ren2013,Huang2023}: $\Delta(\mb{k},T)=\Delta(\mb{k}) \tanh (1.74\sqrt{T_c/T-1})$ where 
$T_c$ is the critical temperature of the superconductor and $\Delta(\mb k, T=0)=\Delta(\mb k)$.

 To find the conductance, we numerically compute all possible scattering probabilities following the scattering matrix formalism (see Appendix~\ref{SM_Sec2} for details). According to the conservation of the probability current, the unitarity relation is always maintained: $\mc{R}^e_\sg + \mc{R}^h_{\sg} + \mc{T}^e_\sg = 1$ where $\mc{T}^e_\sg$ is the probability of electron transmission, which can be finite within the subgap regime only in the presence of finite DOS arising due to BFSs. At zero temperature, the conductance in our N/SC junction is normalized by the normal metal conductance as $G_N = \frac{e^2}{h} \int_{-\pi/2}^{\pi/2} d\theta_e \cos\theta_e \frac{4\cos^2\theta_e}{4\cos^2\theta_e + Z^2}$,
where $Z$ is the dimensionless barrier strength. 

We now present our numerical results at zero temperature for zero and finite value of $\alpha$ in Sec.\,\ref{SubSec:A} and \ref{SubSec:B}, respectively, followed by the finite temperature results 
discussed in Sec.\,\ref{SubSec:C}.
\begin{figure}
	\centering
	\includegraphics[scale=0.28]{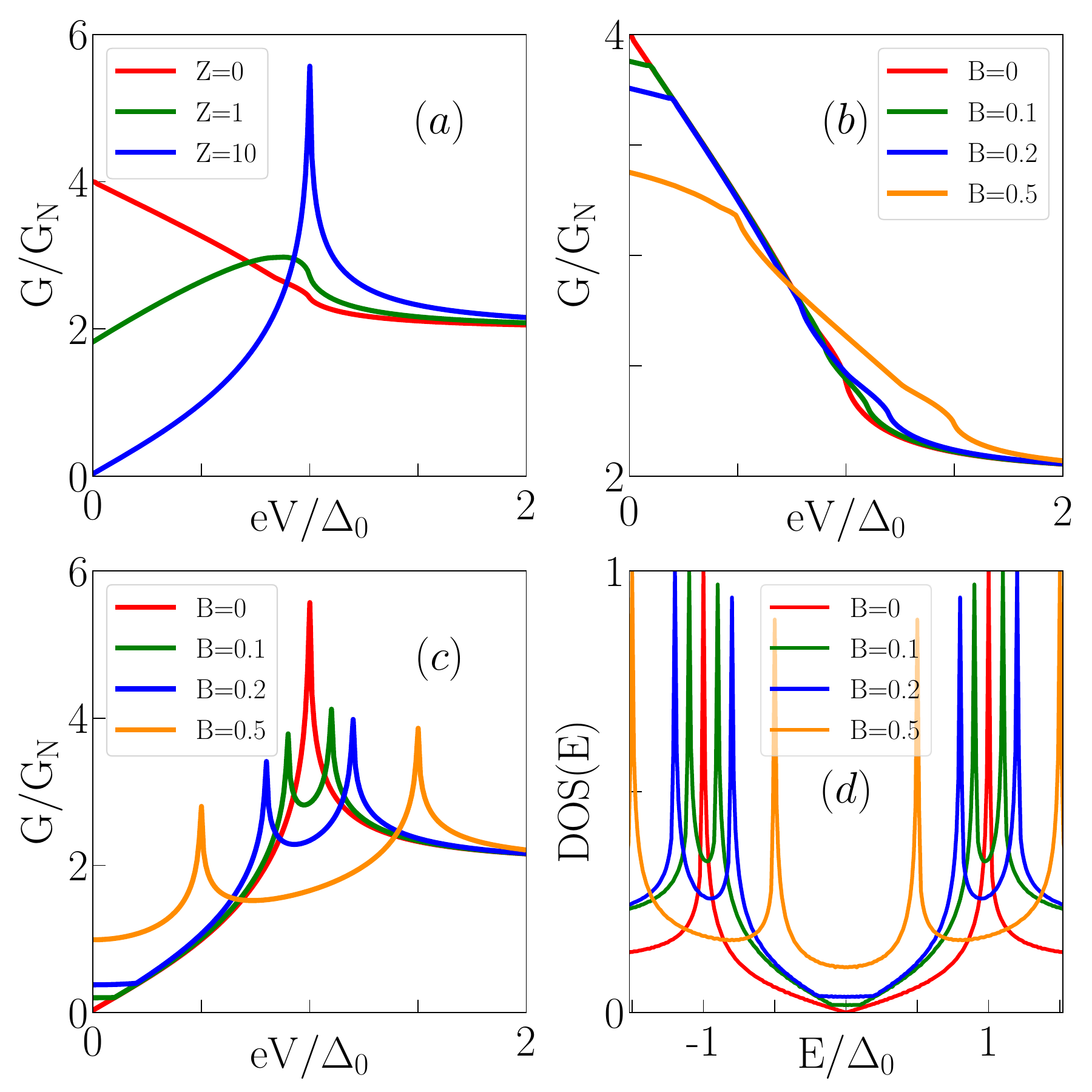} 
\caption{Normalized differential conductance is depicted as a function of voltage bias for $\alpha=0$ in ($a$) absence of Zeeman field ($B=0$), (b) transparent limit ($Z=0$), (c) tunneling limit ($Z=10$). (d) DOS of \dw SC is shown in the presence of Zeeman field $B$. }
\label{fig:Isotropic_dIdV}
\end{figure}

\subsection{$\alpha=0$} \label{SubSec:A}
For the sake of completeness and better understanding, we start with the results in the absence of any Zeeman field, which is already established in the literature\,\cite{Tanaka1995,Perconte2022}. In Fig.~\ref{fig:Isotropic_dIdV}(a), we show the normalized differential conductance $G/G_N$ as a function of $eV$ for
$\alpha=0$ and no magnetic field is applied. In the transparent limit ($Z=0$), we find the decaying nature of the conductance for $eV<\Delta_0$ starting from the value $4$ (considering both spins) at zero energy bias and reach the saturation limit of $2$ for $eV>\Delta_0$. However, the behavior of the conductance qualitatively changes in the presence of a barrier. In the tunnelling limit ($Z\gg 1$), the conductance follows the quasiparticle DOS of the bulk \dw SC with decreasing ZBC from $4$ to $0$ with increasing $Z$. This happens due to the enhancement of NR probability as $Z$ increases. 

At $T=0$, we approximate Eq.\,\eqref{Eq:BTK_formula} for the two transparency limits utilizing the unitarity relation as:
\begin{eqnarray}
G(eV, Z=0)=&\frac{e^2}{h}\sum_\sg \int_{-\pi/2}^{\pi/2} d\theta_e \cos\theta_e (1 + \mc{R}^h)\ ,  \non \\ \!=& \!\frac{e^2}{h}\sum_\sg\int_{-\pi/2}^{\,\pi/2}d\theta_e \cos\theta_e (2-\mc{T}^e)\ , \non \\ \label{dI_dV_Z=0} \\
G(eV, Z\gg 1)\!\sim&\!\frac{e^2}{h} \sum_\sg\int_{-\pi/2}^{\pi/2}d\theta_e \cos\theta_e (1-\mc{R}^e)  \ ,
\! \non \\=&\!\frac{e^2}{h}\sum_\sg\int_{-\pi/2}^{\,\pi/2}d\theta_e \cos\theta_e \mc{T}^e\ .  \non \\ \label{dI_dV_Z=10}
\end{eqnarray}

\begin{figure}
	\centering
	\includegraphics[scale=0.22]{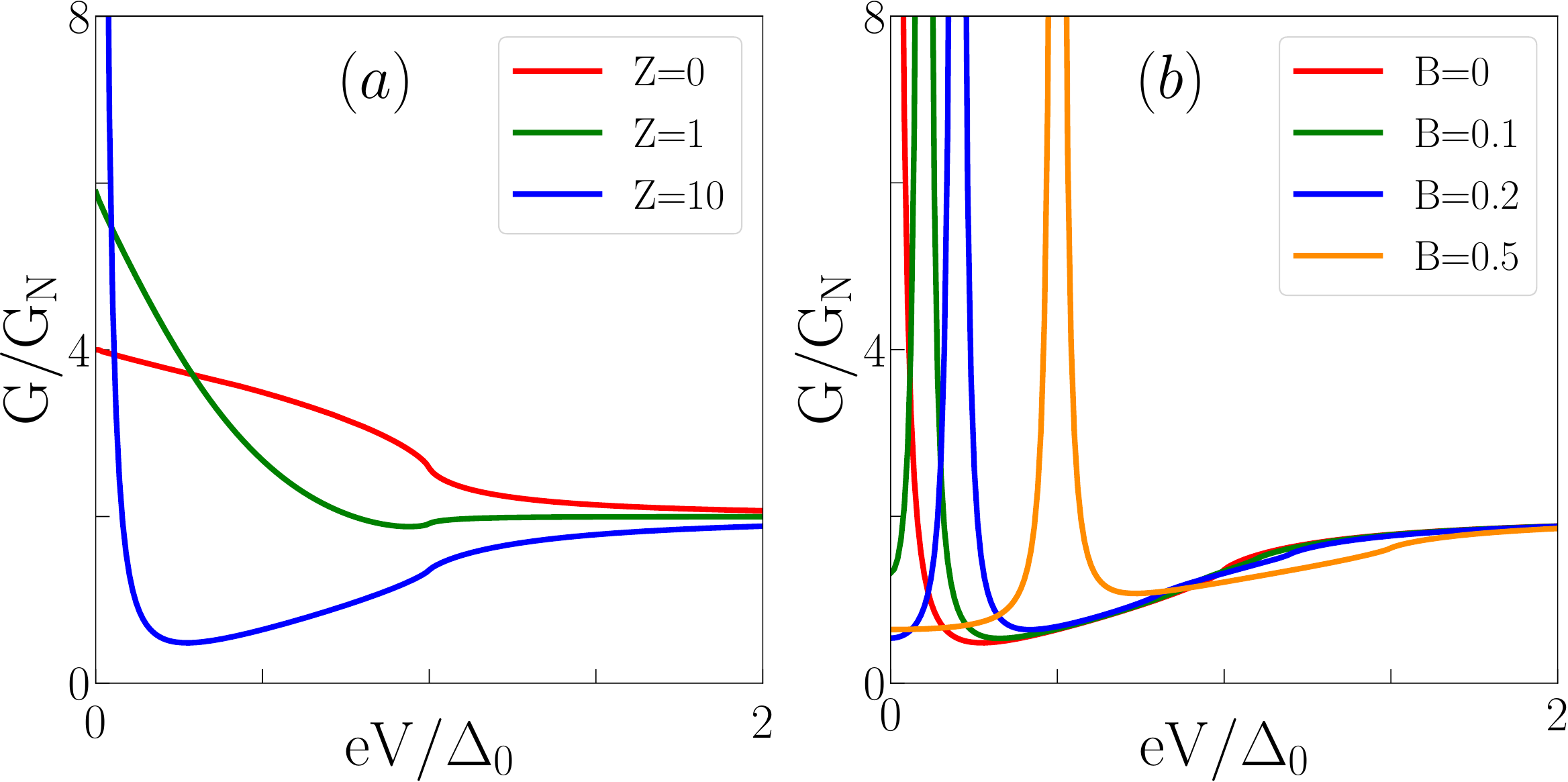} 
	\caption{($a$) Normalized differential conductance is shown as a function of voltage bias for $\alpha=\pi/4$ in (a) absence of Zeeman field ($B=0$) for various barrier strengths, (b) tunneling limit ($Z=10$) for N/\dw SC junction for various values of $B$.} 
	\label{fig:Anisotropic dI_dV}
\end{figure}

Now, we turn on Zeeman field $B$ and depict the corresponding differential conductance in Figs.~\ref{fig:Isotropic_dIdV}(b) and (c) for the two limits of $Z$. In the transparent limit ($Z=0$), we observe that almost flat region appears in the conductance profile and the magnitude of ZBC decreases as we increase $B$. However, when $eV>\Delta_0$, the behavior of appproaching the saturation limit is similar to the previous case. Decrease in conductance for $eV/\Delta_0<1$ means either reduction in the AR probability [see Eq.~(\ref{dI_dV_Z=0})] or equivalently, increase in the transmission 
since NR cannot take place in the fully transparent limit. However, transmission process can take place if the QP DOS is finite for $eV<\Delta_0$, which arises when BFSs accompanied by finite BQP DOS appear. The flat regions around $eV=0$ in the DOS gets elevated gradually with increasing $B$ as shown in Fig.~\ref{fig:Isotropic_dIdV}(d). This clearly reflects in the decreasing behavior of the conductance in the transparent limit (see Fig.~\ref{fig:Isotropic_dIdV}(b)). 

On the other hand, in the tunneling limit ($Z\gg 1$), the conductance peak arising at $eV=\Delta_0$ splits into two and shifts to $eV_\pm=\Delta_0 \pm B$, since the spin degeneracy is lifted as shown in Fig.~\ref{fig:Isotropic_dIdV}(c). This splitting corresponds to the splitting we find in the DOS spectrum as depicted in Fig.~\ref{fig:Isotropic_dIdV}(d). Note that, enhancement of residual DOS at $eV=0$ already indicates the signatures of BFS in presence of the magnetuc field. We calculate the DOS for various parameter values which help in explaining our results. Similar to the $Z=0$ limit, we observe that flat regions appeared in the conductance profile near zero energy. In contrast to $Z=0$ limit, the ZBC increases with magnetic field for $Z\gg 1$. This enhancement of ZBC is due to the generation of BFSs as one increases B [see Fig.~\ref{fig:Isotropic_dIdV}(c)]. BFSs manifest themselves as zero-energy excitations, thus enhance the zero-energy DOS. As a result, the probability of electron transmission through the junction increases. From Eq.\,\eqref{dI_dV_Z=10}, it is evident that this increase in transmission probability enhances the ZBC.


\begin{figure}
	\centering
	\includegraphics[scale=0.22]{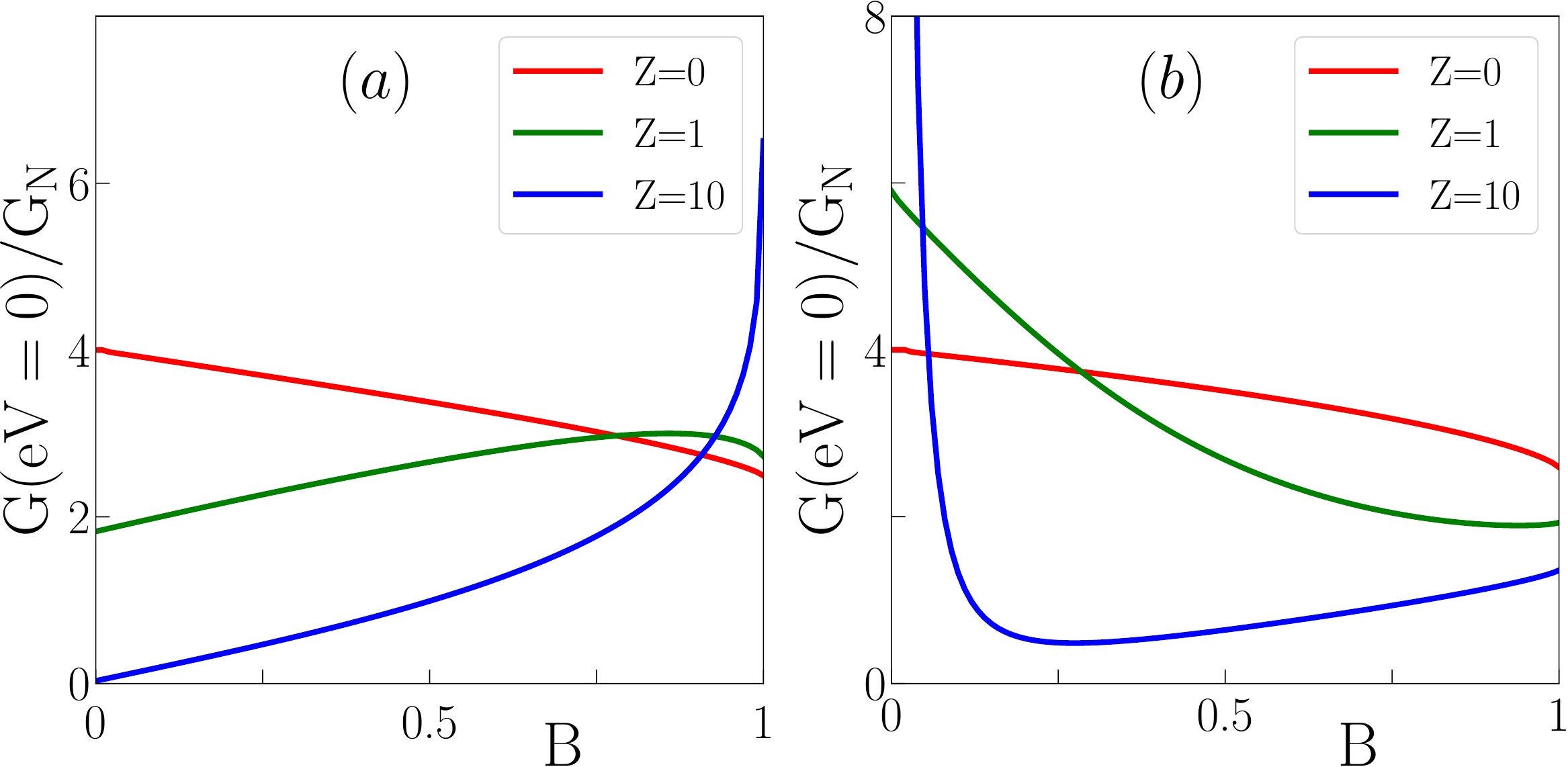} 
	\caption{ZBC is depicted as a function of the magnetic field for (a) $\alpha=0$ and (b) $\alpha=\pi/4$ in N/\dw SC setup shown in Fig.~\ref{fig:Schematic_NS}(a) for various values of barrier strengths.}
	\label{fig:ZBC_vs_B}
\end{figure}

Therefore, in both the transparency limits, we find clear signatures of BFSs. In the literature, some attempts either incorporating disorder\,\cite{Oh2021} or Rashba spin-orbit interaction\,\cite{Banerjee2022}, have been reported to find the signatures of BFSs. The latter one includes $s$-wave SCs~\cite{Banerjee2022}. Here, we show the clear signatures of BFSs in a N/$d$-wave SC junction for the first time which can be possible to realize using a conventional transport measurement setup, and thus enhances the importance of our work.

\subsection{$\alpha=\pi/4$} \label{SubSec:B}
To discuss the transport signature of BFSs in case of nonzero values of $\alpha$, we show $G/G_N$ for $\alpha=\pi/4$ in Fig.\,\ref{fig:Anisotropic dI_dV}(a) and (b). We choose this value of $\alpha$ in order to capture the maximum effect of the anisotropy. For the results at $\alpha=\pi/8$, we refer the readers to Appendix\,\ref{SM_Sec5}. In the transparent limit ($Z=0$), we find the differential conductance resembling the same as that of the $\alpha=0$ case [see Fig.\,\ref{fig:Isotropic_dIdV}(a)]. In contrast, in the tunneling limit ($Z\gg1$), a zero bias conductance peak develops due to the formation of the ABS at the interface of metal/$d$-wave SC junction. For any value of $\alpha$ between $0$ and $\pi/4$ ($0<\alpha\leq \pi/4$), ABSs are formed at the interface due to unconventional nature of pairing potential. It is already established as a signature of the anisotropy in the literature\,\cite{Tanaka1995,Perconte2022}. Interestingly, the ZBC peak is splitted and moves towards the higher $eV$ with increasing $B$ as shown in Fig.\,\ref{fig:Anisotropic dI_dV}(b). Additionally, the ZBC peak developed due to the presence of zero-energy ABS at the interface now splits into two at $E=\pm B$ forming nonzero-energy ABSs resulting in decrease in ZBC. The peak shift is an artifact of adding the magnetic field in the system which breaks the time-reversal symmetry and lifts the spin degeneracy of the system. Similar to the conductance plot for $\alpha=0$ case, where the peak appearing at $E=\Delta_0$ splits into two [see Fig.~\ref{fig:Isotropic_dIdV}(c)], for $\alpha=\pi/4$ as well the zero-bias conductance peak splits and shifted to non-zero energy as shown in Fig.~\ref{fig:Anisotropic dI_dV}(b).
\begin{figure}
	\centering
	\includegraphics[scale=0.22]{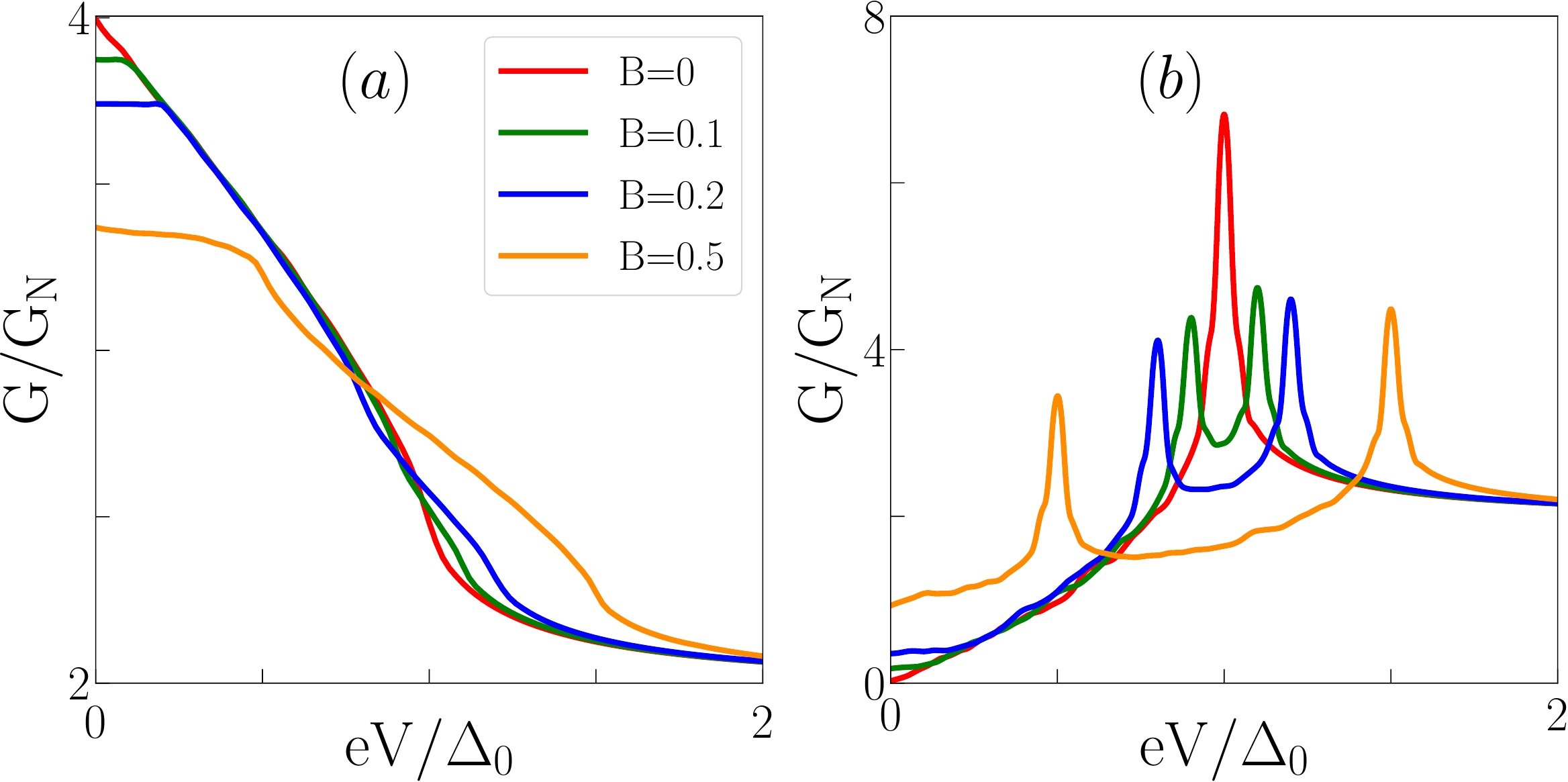} 
	\caption{Normalized differential conductance is shown for the lattice model as a function of voltage bias in ($a$) transparent limit ($Z=0$), (b) tunneling limit ($Z=10$) considering the length and width of the lattice as $100a$ and $500a$ respectively, $a$ being the lattice spacing.}
	\label{Lattice_and_FF}
\end{figure}

To investigate the behavior of the conductance at $eV=0$ in more details, we show the ZBC as a function of $B$ for $\alpha=0$ and $\alpha=\pi/4$ in Figs.\,\ref{fig:ZBC_vs_B}(a) and (b), respectively (see Appendix\,\ref{SM_Sec5} for $\alpha=\pi/8$ results). For $\alpha=0$, we observe a decrease in ZBC with the increase in $B$ in the transparent limit whereas, following the trend of increasing zero-energy DOS with $B$, ZBC rises with $B$ in the tunneling limit as shown in Fig.\,\ref{fig:ZBC_vs_B}(a). Similar feature can be observed in case of $\alpha=\pi/4$ as one increases the value of $B$ [see Fig.\,\ref{fig:ZBC_vs_B}(b)]. Both of these observations can be explained from the generation of BFS as zero-energy excitations in the superconducting side. The states inside BFSs increase the transmission probability of electrons via tunneling through the metal-SC interface. The increase in the transmission probability leads to the reduction of ZBC in the transparent limit [see Eq.\,\eqref{dI_dV_Z=0}] and enhancement of ZBC in tunneling limit [see Eq.\,\eqref{dI_dV_Z=10}]. However, in sharp contrast to $\alpha=0$ case, the ZBC for the $\alpha=\pi/4$ initially falls very fast, but later rises with the increase in $B$ in the tunneling limit as mentioned before. In the transparent limit, the behaviour of ZBC is similar to the $\alpha=0$ case.


We strengthen our results obtained in the continuum model by computing the differential conductance using the python package KWANT~\cite{Groth2014} based on a square lattice (see Appendix~\ref{SM_Sec3} for the discussions on the model). We present the results for the $\alpha=0$ case in Figs.\,\ref{Lattice_and_FF}(a)-(b), both in the transparent and tunnelling limit, which is in good agreement with our continuum model based results [Figs.~\ref{fig:Isotropic_dIdV}(b)-(c)]. Other results for the lattice model are presented and discussed in the 
Appendix~\ref{SM_Sec4}.

\subsection{Effect of finite temperature} \label{SubSec:C}
\begin{figure}
	\centering
	\includegraphics[scale=0.25]{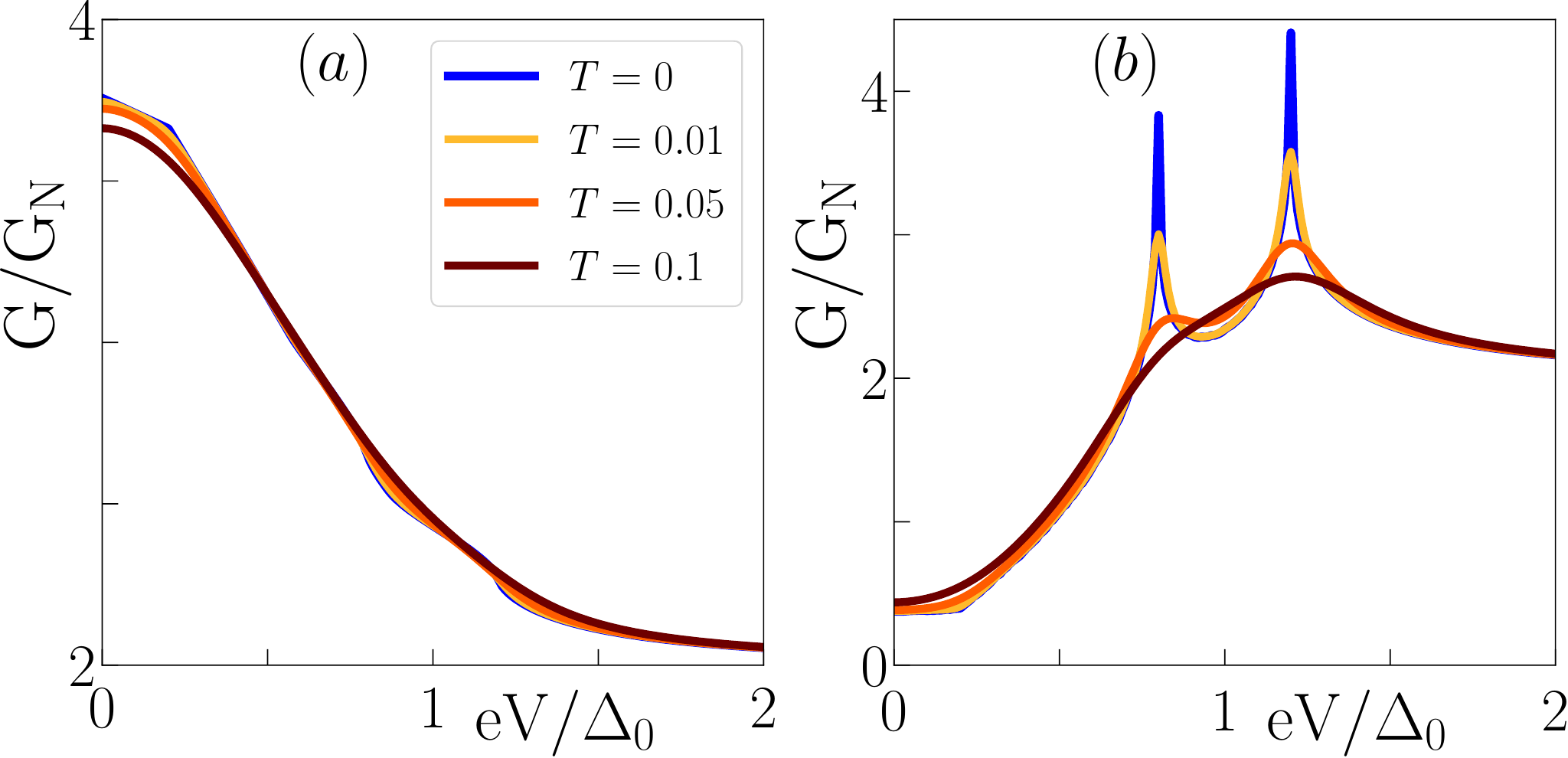} 
	\caption{ Normalized differential conductance is shown as a function of the voltage bias choosing various temperatures (scaled by $T_c$) for $\alpha=0$ and $B=0.2$ in $(a)$ ballistic limit 
	($Z=0$) and $(b)$ tunneling limit ($Z=10$).} 
	\label{fig.dIdV_Tn0}
\end{figure}
In this subsection, we discuss the transport properties of BFS at finite temperature $T$. Formally, the effect of finite temperature enters in Eq.\,\eqref{Eq:BTK_formula} through the Fermi distribution function leading to thermal smearing of the zero-temperature results by the weight factor proportional to $\frac{\partial}{\partial V}\left[ f_L(E,\mu_L,T)  - f_R(E,\mu_R,T)\right] $. Note that, to maintain the superconducting phase, the temperature $T$ should be maintained as less than the superconducting transition temperature $T_c$ ($k_BT_c\approx \Delta_0$). Employing Eq.\,\eqref{Eq:BTK_formula}, 
we compute the normalized differential conductance, $G(eV)/G_N$ at various temperatures and depict the results in Fig.~\ref{fig.dIdV_Tn0} choosing $\alpha = 0$ and $B = 0.2$, in both the ballistic 
($Z = 0$) as well as tunneling ($Z = 10$) limits. As expected, the zero-temperature results get thermally broadened out as we observe in Fig.~\ref{fig.dIdV_Tn0}. Also, at $eV=\Delta$, the quasi-particle 
peak height reduces and gets smeared out due to thermal fluctuations [see Fig.~\ref{fig.dIdV_Tn0}(b)]. We also present the results of ZBC as a function of $B$ in the tunneling limit ($Z = 10$) at finite temperatures in Fig.~\ref{fig.ZBC_Tn0} for both $\alpha = 0$ and $\alpha = \pi/4$. Note that, the infinitesimal increment in conductance at $B=0$ and $\alpha=0$ [see Fig.~\ref{fig.ZBC_Tn0}(a)] emerges due to the thermal population of states in normal metal side when $T\neq 0$. Otherwise, we find that the behavior of the conductance at finite temperatures remains qualitatively similar to that of zero temperature.

\section{Fano factor}\label{sec:IV}
Here, we discuss the behavior of FF to unravel the role of BFS in more detail. Previously, the FF was calculated for the heterotructure considered in Ref.~\cite{Banerjee2022}. Therefore, it can be interesting to investigate the same to identify the signature of BFS in our N/d-wave SC hybrid setup. We find opposite behavior of BFSs in the two different limits of the junction transparency. For $\alpha=0$, ZBC monotonically increases with magnetic field whereas for $\alpha=\pi/4$, ZBC first rapidly decreases then increase 
\begin{figure}[!h]
	\centering
	\includegraphics[scale=0.25]{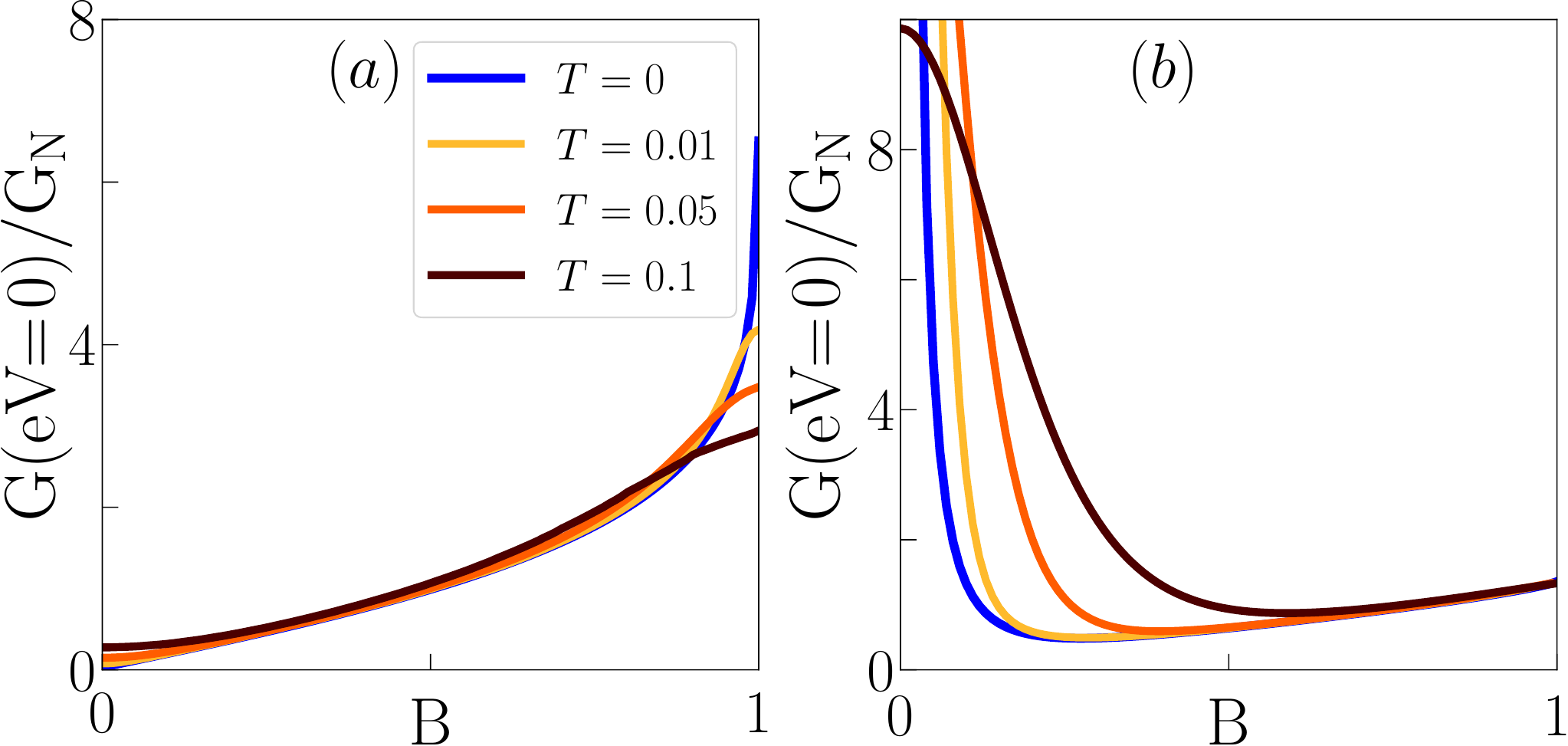} 
	\caption{ZBC is depicted as a function of the magnetic field $B$ at various temperatures (scaled by $T_c$) in the tunneling limit ($Z=10$) for $(a)$ $\alpha=0$ and $(b)$ $\alpha=\pi/4$.} 
	\label{fig.ZBC_Tn0}
\end{figure}
gradually. In order to unravel this significant discrepancy for $\alpha=\pi/4$ as compared to $\alpha=0$,
we investigate the behavior of the FF. We can extract the charge content of the current carriers that tunnel through the junction by analyzing the FF. We predict that the interplay of the two carriers, CPs and BQPs, gives rise to the different intricate features in the conductance spectra. It can be emphasized by examining the nature of the charge carriers participating in the transport. The CPs contribute to the conductance via AR which dominates over other scattering processes in the transparent limit, whereas BQPs contribute via tunneling. Depending on their relative concentration at BFSs we obtain different results for the FF. 

Investigation of noise spectra and computation of FF serve as one of the fundamental tools to explore the nature of charge dynamics in various mesoscopic transport phenomena\,\cite{Anantram1996,Blanter2000,deJong1994,Kobayashi2021,Tanaka2000shotNoise,Zhu1999ShotNoise,Paul2017,Haim2015,Rech2012,Chevallier2011}. FF provides the information about charge content of the current carriers in the weak tunneling limit. Specifically, in metal-SC junctions the charge information of the current carriers that tunnel through the junction can be obtained from the FF in the limit of small transmission probabilities i.e. low junction transparency ($Z\gg 1$) \cite{Anantram1996,Blanter2000,deJong1994,Kobayashi2021,Tanaka2000shotNoise,Zhu1999ShotNoise}. Due to this fact, we demonstrate the FF only in the tunneling limit. 

At zero temperature, the zero-frequency shot noise power for a N/SC junction can be obtained as,
$\mc{S}(eV)\!=\!\int_{0}^{eV} dE~ \mc{S}_T(E)$ with $\dis{\mc{S}_T(E)\!=\!\frac{2e^3}{h}\int_{-\pi/2}^{\pi/2}d\theta \cos\theta ~ \mc{S}_{\theta}(E,\theta)}$ where $\dis{\mc{S}_{\theta}(E,\theta)\!=\![R^e(1-R^e) +  R^h(1- R^h) + 2R^eR^h]}$ \cite{Anantram1996,Blanter2000,Tanaka2000shotNoise,deJong1994,Kobayashi2021,Zhu1999ShotNoise}. The total current is given by, $\dis{I(eV)= \int_{0}^{eV} dE~ G(E)}$ and the FF is the ratio of shot noise power to current given by, $\dis{{\rm{FF}}(eV) = \mc{S}(eV)/2eI(eV)}$~\cite{Blanter2000}. 
\begin{figure}
	\centering
	\includegraphics[scale=0.22]{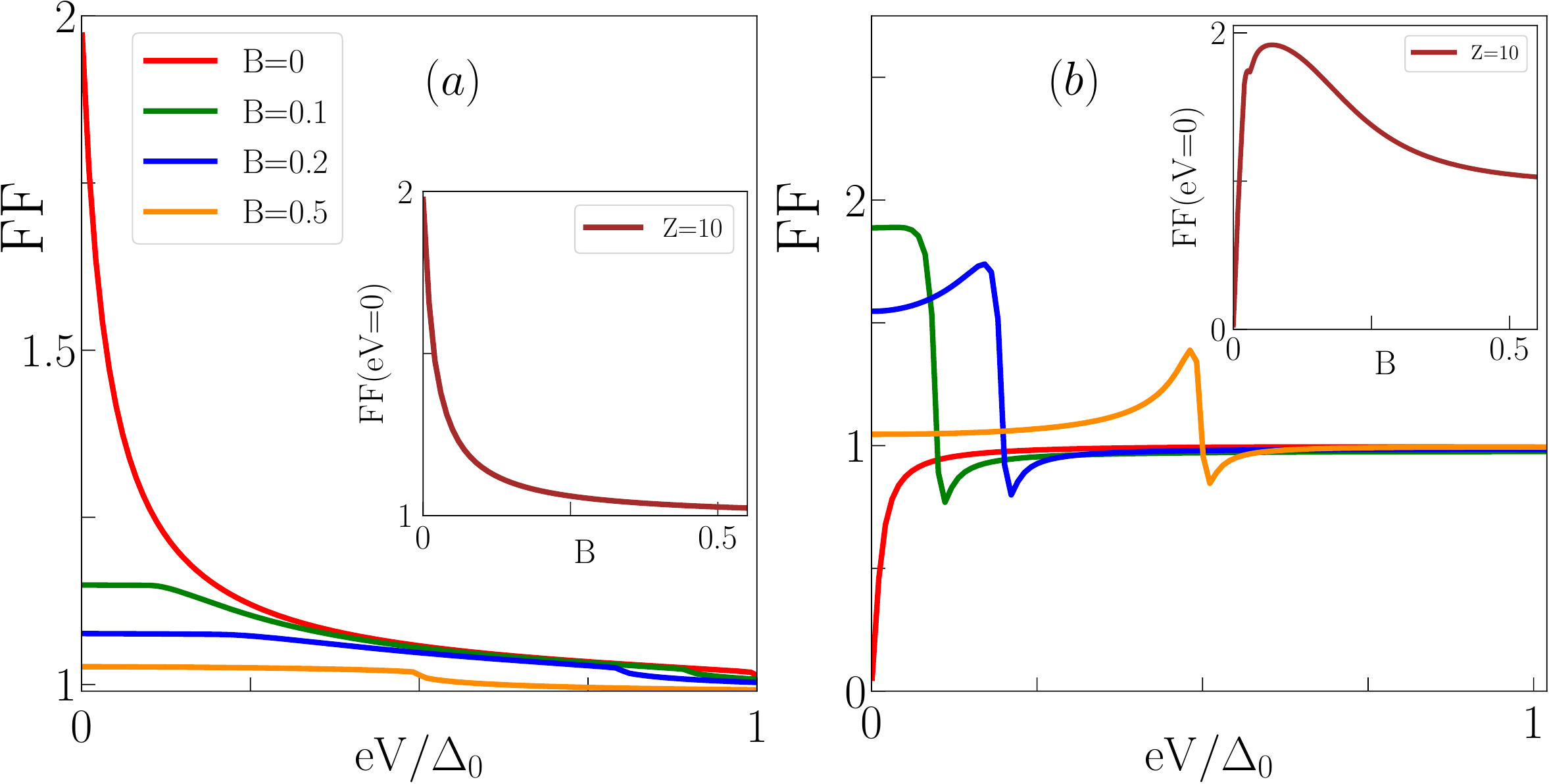} 
	\caption{ FF is illustrated as a function of $eV/\Delta_0$ choosing $Z=10$ for (a) $\alpha=0$ and (b) $\alpha=\pi/4$. 
		Inset: ${\rm{FF}}~(eV=0)$ is highlighted as a function of the Zeeman field $B$. In panels (b) different color lines indicate the same $B$ values as mentioned in panel (a).} 
	\label{Fano factor}
\end{figure}

We numerically compute the FF as a function of the voltage bias for $\alpha$=0 [$\alpha=\pi/4$] and depict the corresponding behavior in Fig.\,\ref{Fano factor}(a) [Fig.~\ref{Fano factor}(b)] considering tunneling limit ($Z\gg 1$). Also, FF ($eV=0$) as a function of $B$ is shown in the insets of these figures. For $\alpha=0$, in the absence of magnetic field, we observe that FF($eV=0$) is 2 indicating the presence of {\it only} CPs at zero energy (see Fig.\,\ref{Fano factor}(a))\,\cite{Tanaka2000shotNoise,Zhu1999ShotNoise}. For $eV>0$, it decreases rapidly and approaches to $1$ due to the nodal lines of \dw SC as established in the literature\,\cite{Tanaka2000shotNoise}. However, for $B\neq0$, we observe a rapid decrease in the FF ($eV=0$) (see Fig.~\ref{Fano factor}(a) inset) due to the generation of BFSs. This indicates the increase in BQPs population inside BFSs which reduces the effective charge of the zero-energy current carriers. Interestingly, this FF ($eV=0$) extends for finite $eV$, giving rise to plateau-like regions in the FF profile confirming the emergence of BFSs (see Fig.\,\ref{Fano factor}(a)). On the other hand, for $\alpha=\pi/4$, in the absence of any $B$, the FF ($eV=0$) is zero due to the presence of zero-energy ABS at $eV=0$ (Fig.\,\ref{Fano factor}(b))\,\cite{Tanaka2000shotNoise,Zhu1999ShotNoise} which effectively suppresses the contribution of CPs, making their charge signatures undetectable. However, for $B\ne0$, ABSs are formed at non-zero energies which manifest itself through sharp drops at $eV=B$ in the FF profile. In the FF ($eV=0$) behavior, there is a sharp rise of the FF followed by a gradual decrease with the increasing $B$ (see Fig.\,\ref{Fano factor}(b) inset). This initial sharp increase, reaching a value close to 2, indicates the presence of both CPs and BQPs for $B\neq 0$, which was previously undetectable due to the zero-energy ABS. As $B$ increases, more BQPs populate the BFSs, leading to a reduction in the effective charge of the zero-energy carriers, and thus resulting in a decrease in the FF ($eV=0$).

Having discussed the features of the FF, we finally address their connection to the behavior of the conductance discussed in the previous section. In the transparent limit ($Z=0$), where $G~(eV=0)$ is dominated by AR, when the Zeeman field is turned on, the relative concentration of BQPs increases inside BFSs, thus reducing the probability of the AR which further lowers the ZBC. Our results of the conductance for the $\alpha=\pi/4$ can be explained in a similar fashion.
In the tunneling limit ($Z\gg1$), AR is suppressed by NR due to the strong barrier potential and $G~(eV)$ effectively depends on the probability of QP transmission. Presence of BFSs increases zero energy DOS which enhances the QP transmission, and hence ZBC increases with $B$ for $\alpha=0$. On the other hand, for $\alpha=\pi/4$, ZBC initially falls sharply since zero-energy ABS shifts to non-zero energy along with a subsequent appearance of non-zero conductance peaks. As BFSs grow in size, more BQPs start populating which increases the tunneling probability, resulting in the enhancement of ZBC. We refer the readers to Appendix\,\ref{SM_Sec5} for the results of FF at other finite $\alpha$ value, \eg~$\alpha=\pi/8$. We observe that the behavior of FF remains qualitatively similar to that 
of $\alpha=\pi/4$. Note that, we restrict ourselves to zero temperature for the calculation of the FF in order to avoid the thermal fluctuations in the system which make the BFSs smeared out, so that we 
can extract the charge information from the FF unambiguously.

\section{Summary and Conclusions}\label{sec:V}
In this article, we propose that the differential conductance measurement at the interface between a normal metal and TRS-broken \dw SC can reveal the signatures of topologically protected BFSs as zero-energy excitations in bulk $d$-wave SC. We have shown an enhancement in ZBC with $B$ as the key signature of BFSs 
for $\alpha=0$. On the contrary, the presence of ZBC peak due to the ABS formed at the interface has been found as an obstruction for the identification of the BFSs in case of $\alpha\ne0$. We resolve this by applying the in-plane Zeeman field and as a result the zero-energy ABS moves to nonzero-energy enabling the detection of BFSs at zero-bias. For better understanding, we further reveal the anomalous behavior (in case of $\alpha=\pi/4$ compared to $\alpha=0$) of the ZBC due to the interplay of BFS and ABS in noise spectroscopy by computing the FF. We further repeat the calculation of the transport coefficients using the lattice model and find remarkable agreement with the continuum model for both $\alpha=0$ and $\alpha=\pi/4$.

Finally, we compare our results with the transport signatures of BFSs found in the junction with conventional ``isotropic gapped'' $s$-wave SCs used in Ref.\,\cite{Banerjee2022}. In their analysis, in-plane magnetic field and intrinsic Rashba SOC are the two basic ingredients to generate the BFS. In contrast, our model don't have any SOC. Also, our $d$-wave gap structure is inherently ``anisotropic and gapless''. Furthermore, in our model, the number of BFSs is fixed to four. Only the size of the BFSs can be controlled via the external magnetic field. However, in Ref.\,\cite{Banerjee2022}, the number and orientation of BFSs can be controlled externally. Moreover,  we have also explored the transport phenomena for various orientation of the crystal axis with respect to the junction normal, quantified by the parameter $\alpha$. This enables us to study the interplay between BFSs and ABS in transport and FF. This degree of freedom is absent in the model of Ref.\,\cite{Banerjee2022}. One final remark 
is in order. The fate of BFS may be affected by the strong correlation present in $d$-wave superconductor like Cuprates and in that case our formalism will not work. This study can be extended by incorporating the strong correlation in the future. Nevertheless, as long as the $d$-wave pairing remains in our system, BFS will appear in presence of a magnetic field and exhibit the transport features 
as mentioned in our article.

\subsection*{Acknowledgments}

A.\,P. acknowledges Debmalya Chakraborty, Pritam Chatterjee, and Arnob Kumar Ghosh for stimulating discussions. A.\,P. and A.\,S.  acknowledge SAMKHYA: High-Performance Computing Facility provided by Institute of Physics, Bhubaneswar, for numerical computations. P.\,D. acknowledges Department of Space (DOS), Government of India and Department of Science and Technology (DST), India (through SERB Start-up Research Grant (File no.\,SRG/2022/001121)) for the financial support.  

\begin{appendix}
\clearpage
\onecolumngrid
\section{Bogoliubov Fermi Surfaces} \label{SM:Sec_1}
In this section, we begin with the band structure and Fermi surfaces (FSs) of the four band continuum model introduced in the main text [see Eq.\,\eqref{Eq:ham}] for both $\alpha=0$ and $\alpha=\pi/4$ and present them in Fig.\,\ref{fig:BFS_isotropic} and Fig.\,\ref{fig:BFS_anisotropic}, respectively. After diagonalizing the Hamiltonian, the energy eigenvalues are obtained as:
\beq
E_{\sg,\eta}(\mb{k})=- \sg B + \eta\sqrt{\eps (\mb{k})^2 + \Del(\mb{k})^2} \ .
\eeq
\begin{figure}[!h]
	\centering
	\includegraphics[scale=0.8]{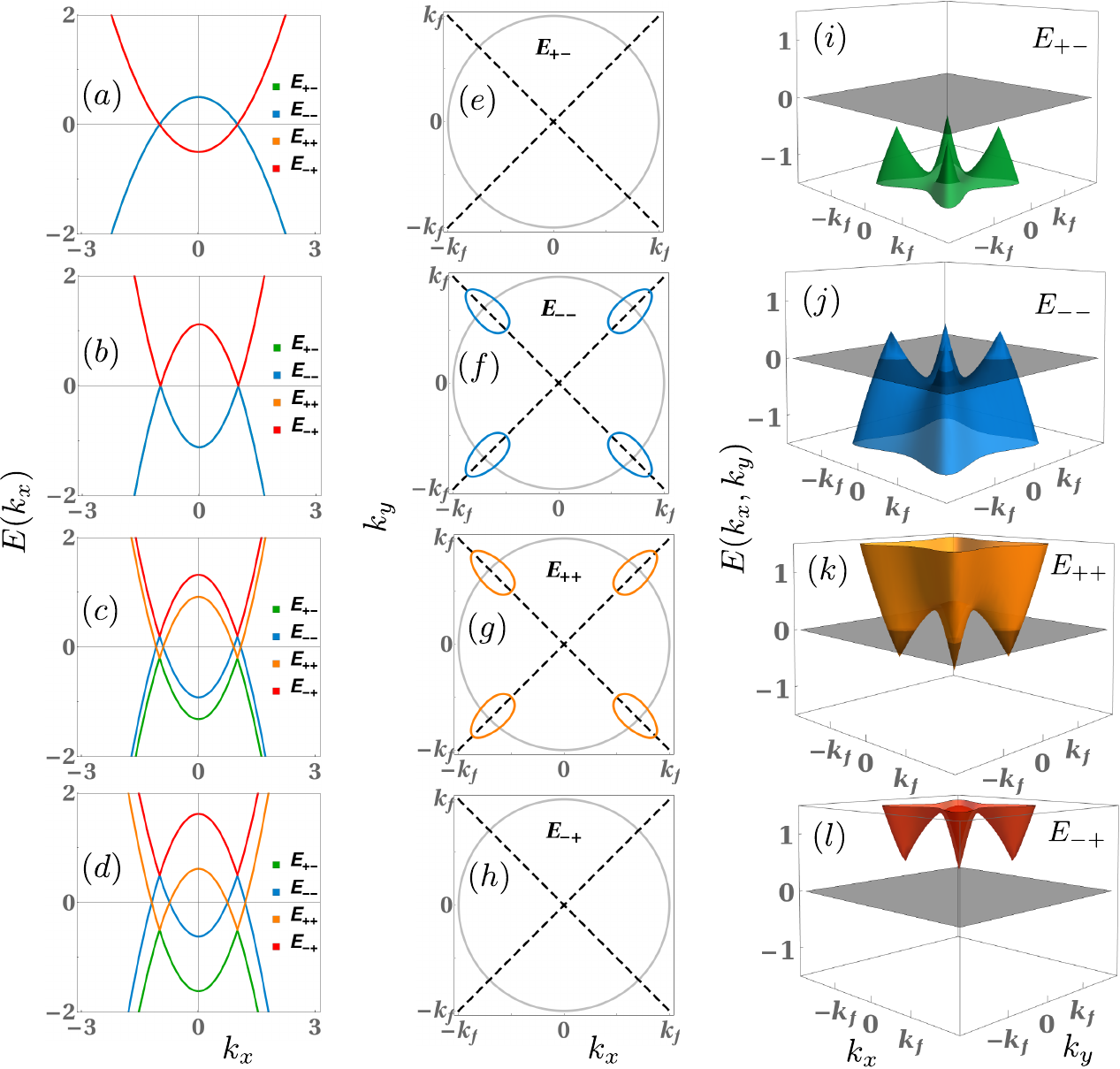}
	\caption{(a)-(d): Energy spectrum is depicted as a function of $k_x$ for the four band model with 
$\alpha=0$ and $k_y=k_f/\sqrt{2}$ where we choose $(\Delta_0,B)={(0,0),(1,0),(1,0.2\Delta_0),(1,0.5\Del_0)}$, respectively. (e)-(h): The Fermi surfaces for the four bands with $(\Del_0,B)=(1,0.5\Del_0)$ are illustrated. BFSs appear in $E_{--} $ and $ E_{++}$. Here, the transparent 
	circle represents the normal state FS. The dotted line denotes the nodal lines of \dw SC. (i)-(l): Band structure in $k_x-k_y$ plane for $(\Del_0,B)=(1,0.5\Del_0)$ is demonstrated for better clarity. 
	Only $E_{--} $ and $ E_{++}$ intersect the zero energy plane (shown by grey color) hosting BFSs. }
	\label{fig:BFS_isotropic}
\end{figure}
\begin{figure}[!h]
	\centering
	\includegraphics[scale=0.8]{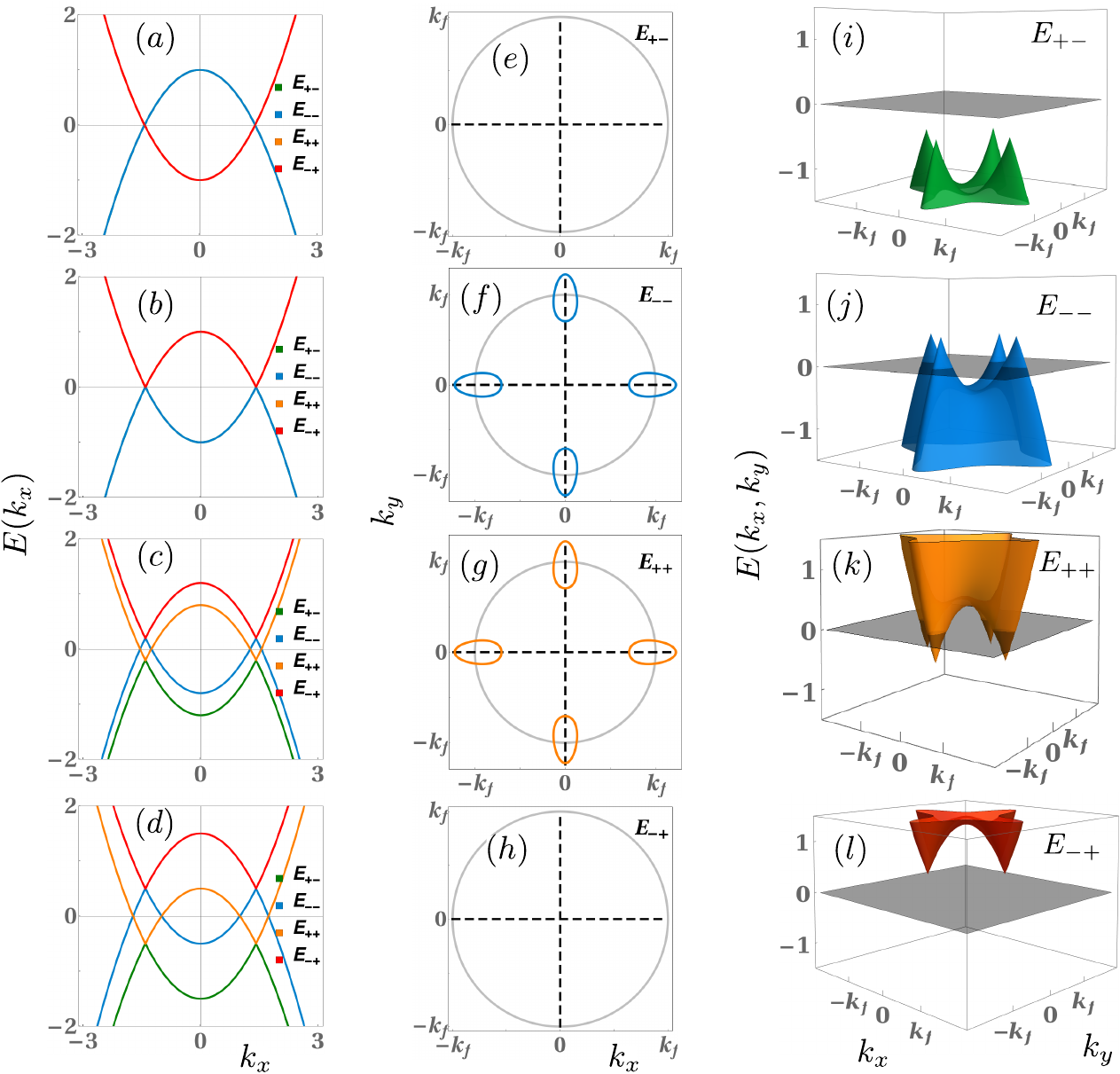}
	\caption{(a)-(d): Energy spectrum is shown as a function of $k_x$ for the four band model with 
$\alpha=\pi/4$ and $k_y=0$ where we choose $(\Delta_0,B)={(0,0),(1,0),(1,0.2\Delta_0),(1,0.5\Del_0)}$, respectively. (e)-(h): Fermi surfaces for the same four bands with $(\Del_0,B)=(1,0.5\Del_0)$ are depicted. BFSs appear in $E_{--} $ and $ E_{++}$ band. The transparent circle represents normal state FS. The dotted lines denote the nodal lines of \dw SC. (i)-(l): The same band structures in $k_x-k_y$ plane for $(\Del_0,B)=(1,0.5\Del_0)$ are shown. Only $E_{--} $ and $ E_{++}$ intersect the zero-energy plane (shown by grey color) hosting BFSs. }	
	\label{fig:BFS_anisotropic}
\end{figure}
For $\alpha=0$, we show the band structure of the four bands in Figs.\,\ref{fig:BFS_isotropic}(a)-(d) considering various values of $(\Del_0,B)$ by fixing $k_y$ at $k_f/\sqrt{2}$ where BFSs appear (the locations of the BFSs are mentioned in the main text). For zero Zeeman field ($B=0$) and SC gap ($\Delta_{0}=0$), the presence of the spin degeneracy leads to doubly degenerate bands in the system, denoted by $E_{++}$, $E_{-+}$, $E_{--}$ and $E_{+-}$. Here, $E_{++}$ and $E_{-+}$ (also $E_{--}$ and $E_{+-}$) overlap with each other as shown in Fig.\,\ref{fig:BFS_isotropic}(a). The conduction bands and valence bands touch each other at $k_x=\pm k_f\sqrt{2}$. With the introduction of the superconductivity, the degeneracy and the band touching phenomena remains the same because of the presence of nodal lines in \dw SC (see Fig.\,\ref{fig:BFS_isotropic}(b)). When the Zeeman field is turned on, the spin degeneracy is lifted and the bands cross each other at multiple points (see Fig.\,\ref{fig:BFS_isotropic}(c-d)). In this scenario, increase in the Zeeman field results in the increase in the splitting of the bands. 

In the absence of any superconductivity and Zeeman field, normal state FS for our four band model is a circle of radius $k_f=\sqrt{2m\mu}$. Introducing \dw superconductivity gives rise to the four nodal lines positioned along $k_x=\pm k_y$. These four nodal lines intersect the normal state FS at four points, $(k_x,k_y)=(\pm k_f/\sqrt{2},\pm k_f/\sqrt{2})$ (see Fig.\, \ref{fig:BFS_isotropic}(e)). In the presence of \dw SC, the normal state FS is gapped out leaving only these four points unaffected. Upon the introduction of a Zeeman field, these point nodes expand, giving rise to the formation of Bogoliubov Fermi surfaces (BFSs) as shown in the main text. However, all the four bands do not host BFSs. Only two bands $E_{++}$ and $E_{--}$ out of the four bands host four BFSs which is evident from Fig.\,\ref{fig:BFS_isotropic}(e-h). We also illustrate the appearance of BFSs in the $k_x-k_y$ plane for all four bands in Fig.\,\ref{fig:BFS_isotropic}(i-l) where only two bands $E_{--}$ and $E_{++}$ intersect the zero energy plane (see Fig.\,\ref{fig:BFS_isotropic}(j-k)) corroborating the emergence of BFSs as shown in Fig.\,\ref{fig:BFS_isotropic}(f-g). The area of intersection increases with the increase in $B$ as shown 
and mentioned in the main text.

For $\alpha=\pi/4$, the nodal lines of \dw SC intersects the normal state FS at $(k_x,~k_y) = (0,\pm k_f)$ and $(\pm k_f,0)$ (see Fig.\,\ref{fig:BFS_anisotropic}(e)). In Fig.\,\ref{fig:BFS_anisotropic} (a-d), we depict all the four bands as a function of $k_x$ with $k_y=0$ and the other parameter values are chosen as the same as $\alpha=0$. In the absence of any Zeeman field, the conduction and valence bands touch each other at $k_x=\pm k_f$. The Fermi surfaces and band structures in $k_x-k_y$ plane for the similar parameter values as for $\alpha=0$ is depicted in Fig.\,\ref{fig:BFS_anisotropic}(e-h) and Fig.\,\ref{fig:BFS_anisotropic}(i-l), respectively. We observe that BFSs appears for the same bands as for the $\alpha=0$ case but rotated by the angle, $\alpha=\pi/4$. This occurs due to the rotation of the nodal lines of the \dw SC. 

In order to characterize the topological property of BFSs, we calculate the Pfaffian for the sake of completeness. Calculation of the Pfaffian relies on the fact that the Hamiltonian can be transformed into an antisymmetric matrix via a unitary transformation \ie
$\tilde{\mc{H}}(\mb{k}) = \Omega\mc{H(\mb{k})}\Omega^\dagger$ so that $ \tilde{\mc{H}}^T(\mb{k}) = -\tilde{\mc{H}}(\mb{k}) $ \cite{Agterberg2017,Brydon2018}. The Hamiltonian in Eq.\,\eqref{Eq:ham} preserves both the charge conjugation ($\mc{C}$) and parity ($\mc{P}$) symmetry individually as well as their product ($\mc{CP}$), and thus satisfies the following relations: 
\bea
\hat{\mc{C}}\mc{H}(\K)\hat{\mc{C}}^{\dagger}& =& -\mc{H}(-\K)\ ,\\ 
\hat{\mc{P}}\mc{H}(\K)\hat{\mc{P}}^{\dagger} &= &\mc{H}(-\K)\ , \\
(\hat{\mc{C}}\hat{\mc{P}})\mc{H}(\K)(\hat{\mc{C}}\hat{\mc{P}})^{\dagger}&=& -\mc{H}(\K)\,.
\eea
 where, 
$\hat{\mc{C}}=U_C\mc{K}=\pi_y\sg_y\mc{K}$ and $\hat{\mc{P}}=U_P=\pi_0\sg_0$ with $\mc{K}$ as complex conjugation operator. Using these expressions, we can rewrite the above relation as, 
 \beq
 U_{CP}\mc{H}^T(\K)U^\dagger_{CP}=-\mc{H}(\K)\ ,
 \eeq
 where $U_{CP}=U_CU^*_P$. For the class of Hamiltonians satisfying this relation, one can construct a unitary matrix $\Omega$ which can transform $\mc{H}(\K)$ into an antisymmetric form $\tilde{\mc{H}}(\mb{k})$. Thus, the Pfaffian $Pf(\K)$, becomes well-defined and takes the form as, $Pf(\K)= \eps(\K)^2 + \Del(\K)^2 - B^2 $ as mentioned in the main text\,\cite{Setty2020PRB}. However, the existence of a well-defined Pfaffian, $Pf(\K)$, does not imply the non-trivial topology of BFSs. For the topological protection of FSs, $Pf(\K)$ must changes its sign over the momentum space. It can be shown that for any non-zero value of $B$, $Pf(\K)$ changes sign in momentum space ensuring the emergence of topological BFSs as mentioned in the main text.

\section{~Scattering Matrix Formalism } \label{SM_Sec2}
Here, we discuss the details of our scattering matrix formulation to compute the differential conductance. We solve the following BdG Hamiltonian to find the quasiparticle excitations having energy $E$ measured with respect to $\mu$, 
\begin{align}
\begin{pmatrix}
H_N({\mb{k}}) -\mu & \Delta_\mb{k}\sigma_0 \\
\Delta_\mb{k}\sigma_0 & \mu - \mathcal{T}H_N(-\mb{k})\mathcal{T}^{-1} \\
\end{pmatrix}
\begin{pmatrix}
u \\
v\\
\end{pmatrix}
&=
E
\begin{pmatrix}
u \\
v \\
\end{pmatrix}\ ,
\end{align}
with $\dis{\Delta_{\mb{k}}=\Delta(\theta)e^{i\phi} \Theta(x)}$ where $\Theta(x)$ is the Heavyside step function and $\phi$ is the global $U(1)$ phase of the SC. $\mathcal{T}=i\sg_y\mathcal{K}$ is the time-reversal operator and $\mc{K}$ denotes the complex conjugation operator as mentioned in the previous section. Here, the Pauli matrix $\mb{\sg}$ acts on the spin space. We consider the normal part Hamiltonian as mentioned in the main text where for the SC ($x>0$) a nonzero pair potential $\Delta(\mb{k})$ is introduced via the proximity-induced effect and it is zero for the normal metal (N) ($x<0$) side. Here, $u$ and $v$ are the two component spinors representing the electron and hole part of the QPs respectively.  A thin insulating barrier with strength $V_0$ is modelled by a $\delta$-function potential at the interface ($x=0$) .

Solving the BdG equation for $x<0$ and $x>0$ separately, we obtain the scattering states as follows. In the normal side ($x<0$), the scattering states are given by, 
\begin{equation*}
\psi^{{\rm{inc}}}_{e,\UP}(x) = \frac{1}{\sqrt{2}}\begin{bmatrix}
~ 1 ~\\ i\\ 0 \\ 0 
\end{bmatrix} e^{ik_e\cos\theta_e x}\ ,
~~\psi^{{\rm{inc}}}_{e,\DN}(x) = \frac{1}{\sqrt{2}}\begin{bmatrix}
~ i ~\\ 1\\ 0 \\ 0 
\end{bmatrix} e^{ik_e\cos\theta_e x}\ ,
\end{equation*} 

\begin{equation*}
\psi^{{\rm{ref}}}_{e,\UP}(x) =\frac{1}{\sqrt{2}} \begin{bmatrix}
~ 1 ~\\ i\\ 0 \\ 0 
\end{bmatrix} e^{-ik_e\cos\theta_e x}\ ,
~~\psi^{{\rm{ref}}}_{e,\DN}(x) =\frac{1}{\sqrt{2}} \begin{bmatrix}
~ i ~\\ 1\\ 0 \\ 0 
\end{bmatrix} e^{-ik_e\cos\theta_e x}\ ,
\end{equation*} 

\begin{equation*}
\psi^{{\rm{ref}}}_{h,\DN}(x) = \frac{1}{\sqrt{2}}\begin{bmatrix}
~ 0 ~\\ 0\\ 1 \\ i 
\end{bmatrix} e^{ik_h\cos\theta_h x}\ ,
~~\psi^{{\rm{ref}}}_{h,\UP}(x) = \frac{1}{\sqrt{2}}\begin{bmatrix}
~ 0 ~\\ 0\\ i \\ 1 
\end{bmatrix} e^{ik_h\cos\theta_h x}\ , 
\end{equation*} 
where $\psi^{{\rm{inc(ref)}}}_{\alpha,\sg}(x)$ denotes the incident (reflected) state with spin $\sg=\{\UP,\DN\}$, and $\alpha=\{e,h\}$ represents the electron and hole states, respectively. The momenta of the incident electron and reflected hole are given by respectively,
\begin{align}
k_e=\sqrt{2m(E+\mu)}\ , \\
k_h=\sqrt{2m(E-\mu)}\ .
\end{align}  
The angle $\theta_e$ and  $\theta_h$ correspond to the electron incident and hole reflection angle, respectively. Using the conservation of the parallel component of the momentum we write,
\beq
k_e\sin\theta_e=k_h\sin\theta_h\ ,
\eeq 
where the expression for $\theta_h$ can be obtained as 
\beq
\displaystyle{\theta_h=\sin^{-1}\left(\sqrt{\frac{\mu + E}{\mu-E}} \sin\theta_e\right)}\ .
\eeq
This condition limits the possibility of Andreev reflection above a critical incident angle given by
\beq
\displaystyle{\theta_h^{{\rm{critical}}}=\sin^{-1}\left(\sqrt{\frac{\mu + E}{\mu-E}}\right)}\ .
\eeq
The scattering states inside the SC~($x>0$) can be written as,

\begin{equation*}
\psi^{{\rm{trans}}}_{eL,\UP}(x) = \frac{1}{\sqrt{2}}\begin{bmatrix}
~ u_\UP(\theta_+) ~\\ i\,u_\UP(\theta_+)\\i\, v_\UP(\theta_{+})e^{-i\phi_{+}} \\ v_\UP(\theta_{+})e^{-i\phi_{+}} 
\end{bmatrix} e^{ik_{eL,\UP}x\cos\theta_{eL,\UP} }\ ,
%
%
~~\psi^{{\rm{trans}}}_{eL,\DN}(x) = \frac{1}{\sqrt{2}}\begin{bmatrix}
~i\, u_\DN(\theta_+)~\\u_\DN(\theta_+) \\ v_\DN(\theta_+)e^{-i\phi_+}\\ i\,v_\DN(\theta_+)e^{-i\phi_+} 
\end{bmatrix} e^{ik_{eL,\DN}x\cos\theta_{eL,\DN} }\ ,
\end{equation*} 

\begin{equation*}
\psi^{{\rm{trans}}}_{hL,\UP}(x) =\frac{1}{\sqrt{2}} \begin{bmatrix}
~ v_\UP(\theta_-) \\ i\,v_\UP(\theta_-)\\i\, u_\UP(\theta_{-})e^{-i\phi_{-}} \\  u_\UP(\theta_{-})e^{-i\phi_{-}} 
\end{bmatrix} e^{-ik_{hL,\UP}x\cos\theta_{hL,\UP} }\ ,
%
%
~~\psi^{{\rm{trans}}}_{hL,\DN}(x) = \frac{1}{\sqrt{2}}\begin{bmatrix}
~ i\,v_\DN(\theta_-)\\v_\DN(\theta_-) \\  u_\DN(\theta_-)e^{-i\phi_-} \\ i\,u_\DN(\theta_-)e^{-i\phi_-} 
\end{bmatrix} e^{-ik_{hL,\DN}x\cos\theta_{hL,\DN} }\ ,
\end{equation*}
where $\psi^{{\rm{trans}}}_{eL(hL),\sg}$ represents the transmitted electron-like (hole-like) quasi-particle state with spin $\sg$. The momenta of the transmitted electron and hole like states 
with spin $\sg$ are respectively given by,
\begin{subequations}
\begin{align}
k_{eL,\sg}=\sqrt{2m(\mu +\sqrt{(E-\sg B)^2-|\Delta(\theta_+)|^2})}\ , \\
k_{hL,\sg}=\sqrt{2m(\mu -\sqrt{(E-\sg B)^2-|\Delta(\theta_-)|^2})}\ .
\end{align}
\end{subequations}
The superconducting coherence factors are given as follow:
\begin{subequations}
\begin{align}
\dis{u_\sg(\theta_\pm)=\frac{1}{\sqrt{2}} \left[ 1 + \sqrt{\frac{(E-\sg B)^2 - |\Delta(\theta_\pm)|^2}{(E-\sg B)^2}}\right]^{1/2}}\ , \\
\dis{v_\sg(\theta_\pm)=\frac{1}{\sqrt{2}} \left[ 1 - \sqrt{\frac{(E-\sg B)^2 - |\Delta(\theta_\pm)|^2}{(E-\sg B)^2}}\right]^{1/2}}\ .
\end{align}
\end{subequations}

For $\alpha\neq 0$, the electron (hole)-like QPSs experience the pair potential as,
\beq
\Delta(\theta_\pm)=\Delta_0\cos(2\theta\mp2\alpha)\ ,
\eeq
satisfying the relation $\dis{e^{i\phi_\pm}=e^{i\phi} \frac{\Delta(\theta_\pm)}{|\Delta(\theta\pm)|}}$.
The transmission angles for the electrons and holes inside the SC can be obtained employing the conservation of the $y$-component of wave-vector i.e. 
\bea
k_{eL,\sg}\sin\theta_{eL,\sg}&=&k_{e}\sin\theta_e\ , \\
k_{hL,\sg}\sin\theta_{hL,\sg}&=&k_{e}\sin\theta_e\ ,
\eea 
where,
\begin{subequations}
\begin{align}
\dis{\theta_{eL,\sg}=\sin^{-1}\left(\sqrt{\frac{\mu + E}{\mu + \sqrt{(E-\sg B)^2 -|\Delta(\theta_+)|^2}}} \sin\theta_e \right)}\ , \\
\dis{\theta_{hL,\sg}=\sin^{-1}\left(\sqrt{\frac{\mu + E}{\mu - \sqrt{(E-\sg B)^2 -|\Delta(\theta_-)|^2}}} \sin\theta_e \right)}\ .
\end{align}
\end{subequations}
Having the scattering states in both sides of the interface, we can write the wave functions in these two regions as,
\begin{eqnarray}
\Psi^N_\sg(x) &=& \psi^{{\rm{inc}}}_{e,\sg} + r^{ee}_{\sg}\, \psi^{{\rm{ref}}}_{e,\sg} +\, r^{eh}_{\bar{\sg}}\, \psi^{{\rm{ref}}}_{h,\bar{\sg}}  , \\
\Psi^{SC}_\sg (x) &=& t^{ee}_{\sg}\,\psi^{{\rm{trans}}}_{eL,\sg} + t^{eh}_{\sg}\,\psi^{{\rm{trans}}}_{hL,\sg},
\end{eqnarray}
where $r^{ee}_{\sg},r^{eh}_{\sg},t^{ee}_{\sg},t^{eh}_{\sg}$ denote the scattering amplitudes for the NR, AR, electron-like, and hole-like transmission, respectively. We obtain these amplitudes using boundary conditions given by,
\begin{subequations}
\begin{align}
\Psi^N_\sg(x=0) &= \Psi^{SC}_\sg(x=0) \label{NS}\ ,\\
\frac{d}{dx}\Psi^{SC}_\sg(x) \vline_{_{x=0^+}}-\frac{d}{dx}\Psi^{N}_\sg(x) \vline_{_{x=0^-}}  &= Z\Psi^N_\sg(x=0) \ .
\label{NS_deriv}
\end{align}
\end{subequations}
Here, $Z=\frac{2mV_0}{\hbar^2k_f}$ is the dimensionless barrier strength. We numerically solve for $r^{ee}_{\sg},r^{eh}_{\sg},t^{ee}_{\sg},t^{eh}_{\sg}$ using Eqs.\,\eqref{NS} $\&$ \eqref{NS_deriv} to compute the differential conductance and shot-noise spectra within the BTK formalism.

\section{~Differential conductance and fano factor for $\alpha=\pi/8$}  \label{SM_Sec5} 
\begin{figure}[h!]
	\includegraphics[scale=0.6]{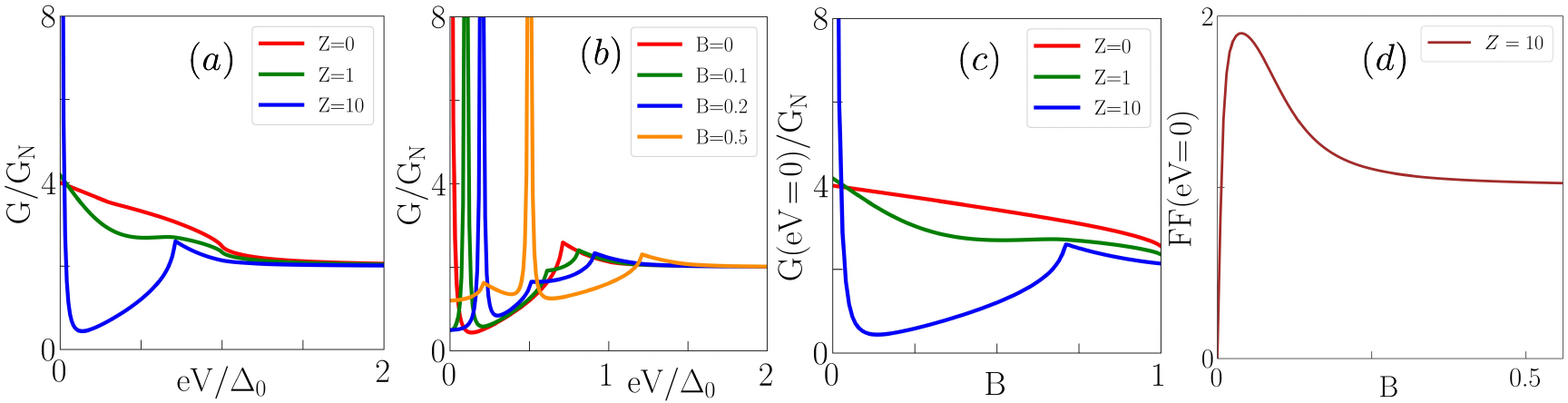} 
	\caption{Normalized differential conductance is shown as a function of voltage bias for $\alpha=\pi/8$ in $(a)$ absence of Zeeman field $(B = 0)$ choosing various barrier strengths, $(b)$ 
	tunneling limit $(Z = 10)$ for different values of the Zeeman field, $B$. $(c)$ Zero bias conductance is depicted as a function of $B$ for various values of the barrier strengths. ($d$) FF~$(eV=0)$ 
	is shown as a function of the Zeeman field in the tunneling limit.}
	\label{Fig.alpha_pi_8}
\end{figure}
In this section, we present the results for the differential conductance and fano factor considering the angle $\alpha=\pi/8$ at zero temperature. We show the normalized differential conductance, $G(eV)/G_N$, as a function of voltage bias $eV$ in Fig.~\ref{Fig.alpha_pi_8}($a$)-($b$), ZBC as a function of magnetic field in Fig.~\ref{Fig.alpha_pi_8}($c$), and zero-bias FF as a function of magnetic field in the tunneling limit ($Z=10$) in Fig.~\ref{Fig.alpha_pi_8}$(d)$. We observe that for large barrier strength, the ABS appears at zero energy in the absence of magnetic field [see Fig.\ref{Fig.alpha_pi_8}($a$)]. Application of an external in-plane magnetic field results in shifting of the ABS to the positive energy [see Fig.~\ref{Fig.alpha_pi_8}($b$)]. Overall, for $\alpha=\pi/8$, the results are qualitatively similar to 
the results for $\alpha=\pi/4$ [see Fig.~\ref{fig:Anisotropic dI_dV}(b)]. In Fig.~\ref{Fig.alpha_pi_8}($d$), we also depict the behaviour of the FF for $\alpha=\pi/8$. This remains qualitatively similar to that of 
$\alpha=\pi/4$ [see the inset of Fig.~\ref{Fano factor}($b$)]. Thus, we infer that for any non-zero value of $\alpha$, ABS is present in the system and it gets shifted to the positive energy with the application of the magnetic field. Further, the corresponding behavior of the differential conductance is expected to remain similar as $\alpha=\pi/4$ even at finite temparature.


\section{~Lattice Model}  \label{SM_Sec3} 

In this section, we construct the lattice model starting from the continuum model introduced in the main text.  We choose a square lattice with the lattice spacing $a=1$. The lattice Hamiltonian can be 
found by replacing $k_i^2 \simeq 2(1-\cos k_i), k_i \simeq \sin k_i$ where $i=x,y$ as,
\bea
\mc{H}_{{\rm{lat}}}(\mb{k}) =[-2t(\cos k_x + \cos k_y) + (4t-\mu)]  \pi_z  \sg_0 -B \pi_0  \sg_y + \Del_{{\rm{lat}}}(\mb{k},\alpha) \pi_x \sg_0
 \label{lattice_model}
 \eea
 where, $\Del_{{\rm{lat}}}(\mb{k},\alpha=0)=2\Del_0(\cos k_y - \cos k_x)$ and $\Del_{{\rm{lat}}}(\mb{k},\alpha=\pi/4)= 2\Delta_0\sin k_x\sin k_y$.  We illustrate schematically the \dw pair potential in the real space on a square lattice schematically for both the $\alpha=0$ [see Fig.~\ref{fig:Schematic_RS}(a)] and $\alpha=\pi/4$ case [see Fig.\ref{fig:Schematic_RS}(b)]. For $\alpha=0$, the pairing happens between two nearest-neighbour sites with opposite spins whereas, for $\alpha=\pi/4$, the pairing is considered between the next-nearest-neighbours with opposite spins.
\begin{figure}[!h]
	\centering
	\includegraphics[scale=0.28]{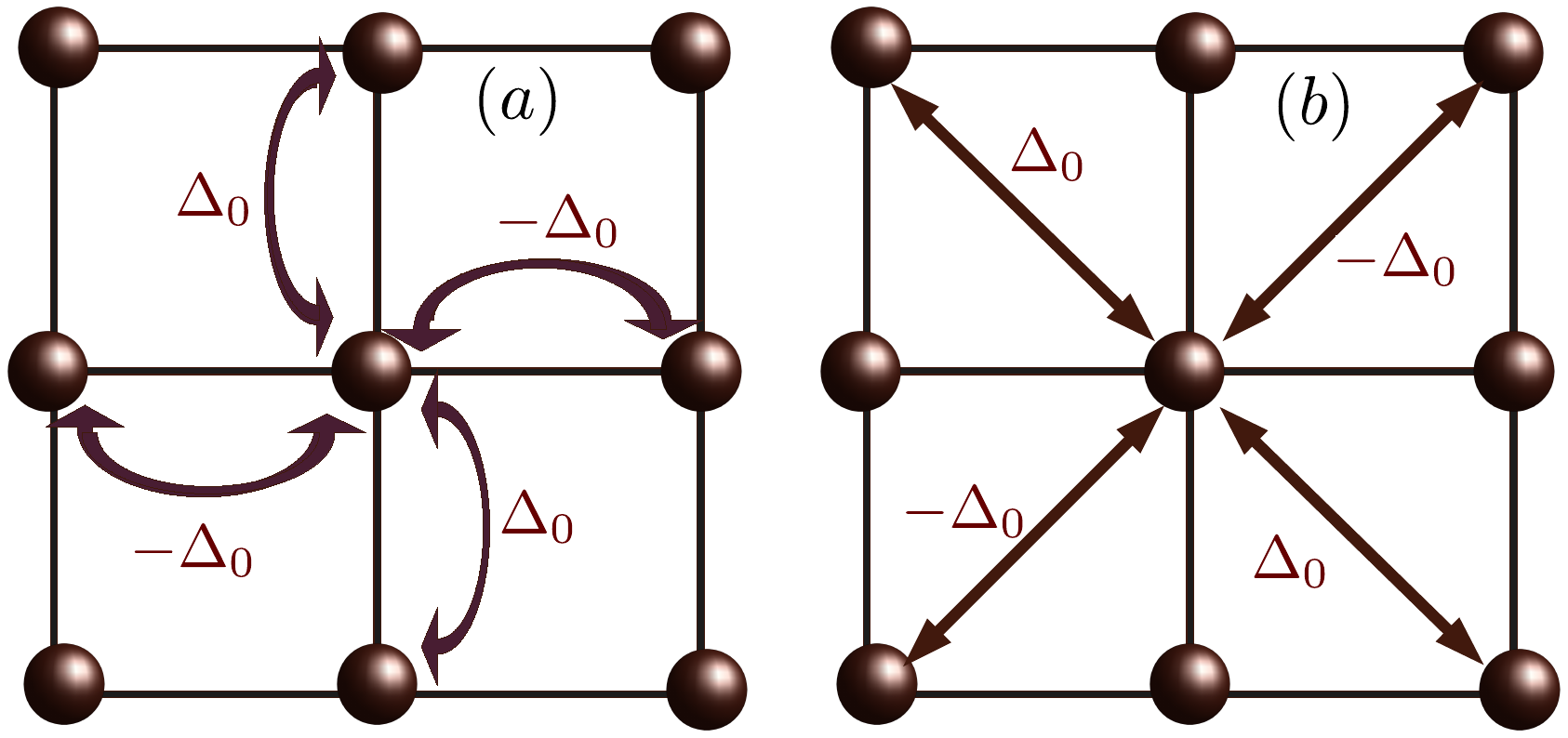}
  	\caption{Schematic diagram of our lattice model with the \dw pair amplitudes in the real space. (a) 
For $\alpha=0$, two nearest-neighbour electrons with opposite spins are paired (denoted by the curved arrows), whereas (b) for $\alpha=\pi/4$, two next-nearest-neighbour electrons are paired (indicated by the straight arrows).}
	\label{fig:Schematic_RS}
\end{figure}
 
\section{~Differential conductance in case of lattice model}  \label{SM_Sec4} 
We also compute the normalized differential conductance for the lattice model using the python package KWANT for a few more parameter values\,\cite{Groth2014}. In the main text, the differential conductance for the N/\dw SC hybrid junction is presented based on our lattice model. Here, we present the normalized differential conductance considering the lattice model for the $\alpha=\pi/4$ case. The corresponding behavior for both the transperant (ballistic) and tunneling limit is depicted in Fig.~\ref{fig:Lattice_aniso}(a) and Fig.~\ref{fig:Lattice_aniso}(b), respectively. For this purpose, we consider a square lattice with dimension $N_x\times N_y$ and attach leads along $x$-direction at $x=\pm N_x$. The leads are modelled using the same Hamiltonian mentioned in Eq.\,\eqref{lattice_model} with $B=0,~\Delta_{0}=0$. We choose the following parameters: $t=1,\mu=t,\Delta_0=0.01\mu, N_x=100a, Ny=350a$ ($a=1$) where $B$ is measured in units of $\Del_0$. We find a fantastic agreement (in both transperant and tunneling limit) with the results obtained in case of continuum model for the same set of parameter values. This agreement enhances the potential of our work.
\begin{figure}[!h]
	\includegraphics[scale=0.25]{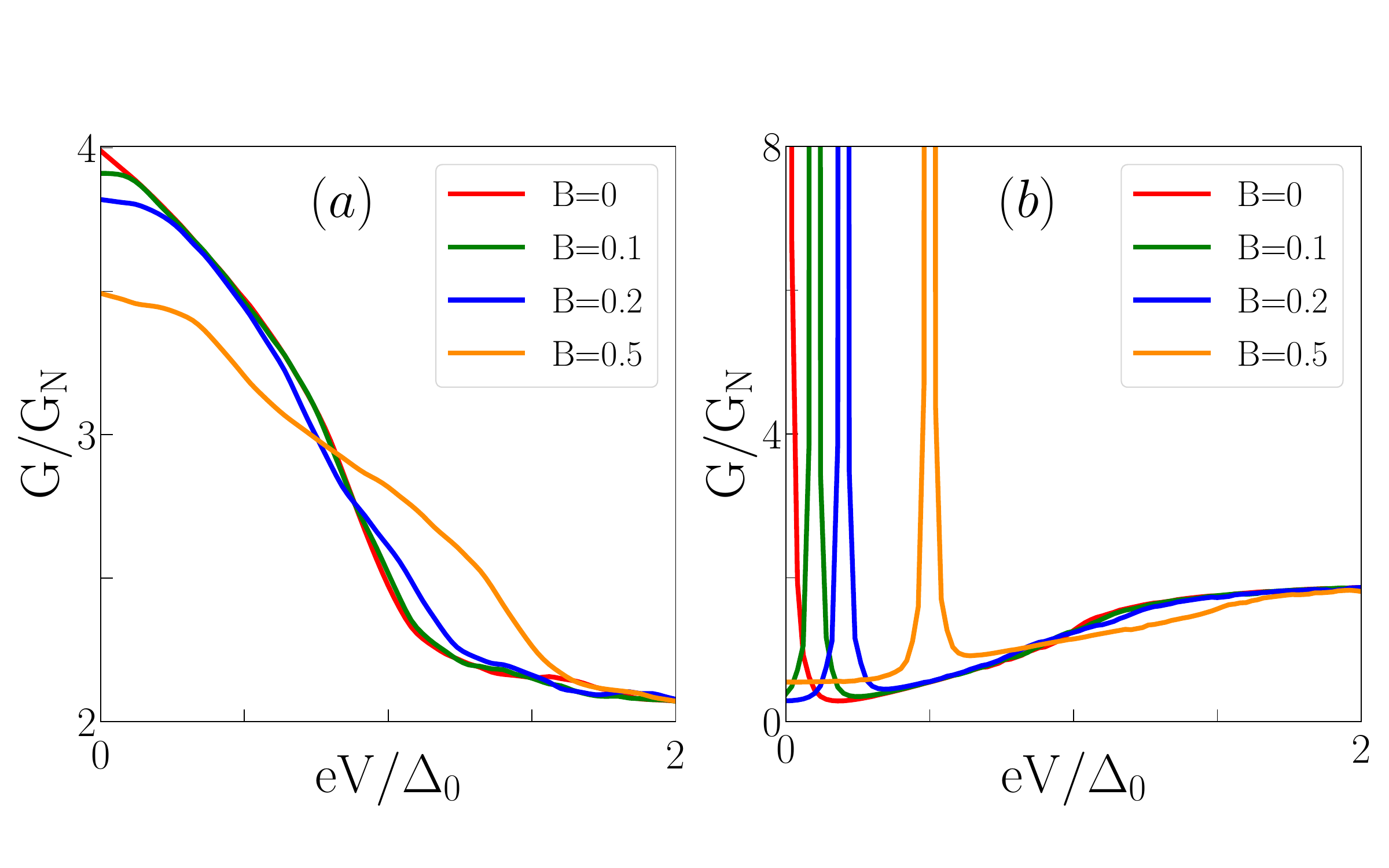} 
	\caption{Normalized differential conductance is depicted considering the lattice model of 
N/\dw SC junction for $\alpha=\pi/4$, as a function of the voltage bias choosing various values of the Zeeman field $B$ in the (a) ballistic ($Z=0$) and (b) tunneling ($Z=10$) limit.}
	\label{fig:Lattice_aniso}
\end{figure}

\end{appendix}
\twocolumngrid
\bibliography{bibfile}
\end{document}